\documentclass[superscriptaddress,aps,preprintnumbers,amsmath,amssymb,prd,nofootinbib,preprint]{revtex4-1}
\pdfoutput=1

\usepackage{graphicx}
\usepackage{epstopdf}
\usepackage{dcolumn}
\usepackage{bm}
\usepackage{hyperref}
\usepackage{color}
\usepackage{amsmath}
\usepackage{multirow}

\begin{document}


\def\a{\alpha}
\def\b{\beta}
\def\c{\varepsilon}
\def\d{\delta}
\def\e{\epsilon}
\def\f{\phi}
\def\g{\gamma}
\def\h{\theta}
\def\k{\kappa}
\def\l{\lambda}
\def\m{\mu}
\def\n{\nu}
\def\p{\psi}
\def\q{\partial}
\def\r{\rho}
\def\s{\sigma}
\def\t{\tau}
\def\u{\upsilon}
\def\v{\varphi}
\def\w{\omega}
\def\x{\xi}
\def\y{\eta}
\def\z{\zeta}
\def\D{\Delta}
\def\G{\Gamma}
\def\H{\Theta}
\def\L{\Lambda}
\def\F{\Phi}
\def\P{\Psi}
\def\S{\Sigma}

\def\o{\over}
\def\beq{\begin{align}}
\def\eeq{\end{align}}
\newcommand{\gsim}{ \mathop{}_{\textstyle \sim}^{\textstyle >} }
\newcommand{\lsim}{ \mathop{}_{\textstyle \sim}^{\textstyle <} }
\newcommand{\vev}[1]{ \left\langle {#1} \right\rangle }
\newcommand{\bra}[1]{ \langle {#1} | }
\newcommand{\ket}[1]{ | {#1} \rangle }
\newcommand{\EV}{ {\rm eV} }
\newcommand{\KEV}{ {\rm keV} }
\newcommand{\MEV}{ {\rm MeV} }
\newcommand{\GEV}{ {\rm GeV} }
\newcommand{\TEV}{ {\rm TeV} }
\newcommand{\1}{\mbox{1}\hspace{-0.25em}\mbox{l}}
\newcommand{\headline}[1]{\noindent{\bf #1}}
\def\diag{\mathop{\rm diag}\nolimits}
\def\Spin{\mathop{\rm Spin}}
\def\SO{\mathop{\rm SO}}
\def\O{\mathop{\rm O}}
\def\SU{\mathop{\rm SU}}
\def\U{\mathop{\rm U}}
\def\Sp{\mathop{\rm Sp}}
\def\SL{\mathop{\rm SL}}
\def\tr{\mathop{\rm tr}}
\def\mpl{M_{\rm Pl}}

\def\IJMP{Int.~J.~Mod.~Phys. }
\def\MPL{Mod.~Phys.~Lett. }
\def\NP{Nucl.~Phys. }
\def\PL{Phys.~Lett. }
\def\PR{Phys.~Rev. }
\def\PRL{Phys.~Rev.~Lett. }
\def\PTP{Prog.~Theor.~Phys. }
\def\ZP{Z.~Phys. }

\def\dd{\mathrm{d}}
\def\ff{\mathrm{f}}
\def\BH{{\rm BH}}
\def\inf{{\rm inf}}
\def\ev{{\rm evap}}
\def\eq{{\rm eq}}
\def\SM{{\rm sm}}
\def\Mpl{M_{\rm Pl}}
\def\GeV{{\rm GeV}}
\newcommand{\Red}[1]{\textcolor{red}{#1}}
\newcommand{\TL}[1]{\textcolor{blue}{\bf TL: #1}}


\title{
Hubble induced mass after inflation  in spectator field models
}

 \author{Tomohiro Fujita}
\affiliation{Stanford Institute for Theoretical Physics and Department of Physics, Stanford University, Stanford, CA 94306, USA}

\author{Keisuke Harigaya}
\affiliation{Department of Physics, 
 University of California, Berkeley, California 94720, USA}
\affiliation{Theoretical Physics Group, 
 Lawrence Berkeley National Laboratory, Berkeley, California 94720, USA}

\begin{abstract}
Spectator field models
such as the curvaton scenario and the modulated reheating
 are attractive scenarios for the generation of the cosmic curvature perturbation, as the constraints on inflation models are relaxed.
In this paper, we discuss the effect of Hubble induced masses on the dynamics of spectator fields after inflation.
We pay particular attention to the Hubble induced mass by the kinetic energy of an oscillating inflaton, which is generically unsuppressed but often overlooked.
In the curvaton scenario, the Hubble induced mass relaxes the constraint on the property of the inflaton and the curvaton, such as the reheating temperature and the inflation scale.
We comment on the implication of our discussion for baryogenesis in the curvaton scenario.
In the modulated reheating, the predictions of models e.g.~the non-gaussianity can be considerably altered. Furthermore,
we propose a new model of the modulated reheating utilizing the Hubble induced mass
which realizes a wide range of the local non-gaussianity parameter.
\end{abstract}


\maketitle

\section{Introduction}

The large scale structure of the universe and the fluctuation of the cosmic microwave background (CMB) are naturally explained by the perturbation generated during cosmic inflation~\cite{Mukhanov:1981xt,Hawking:1982cz,Starobinsky:1982ee,Guth:1982ec,Bardeen:1983qw}.
In the simplest scenario, a scalar field called an inflaton is responsible both for the inflation and the generation of the cosmic perturbation.

In general,
any light scalar field obtains a fluctuation of its field value during inflation and can be the origin of the cosmic perturbation,
even if their energy density is negligible during inflation.
We refer to such field as a spectator field.
In the curvaton scenario~\cite{Mollerach:1989hu,Linde:1996gt,Enqvist:2001zp,Lyth:2001nq,Moroi:2001ct}, the fluctuation turns into a fluctuation of the energy density of the universe as the spectator field dominates the energy density of the universe.
In the modulated reheating~\cite{Dvali:2003em,Kofman:2003nx}, the decay rate of the inflaton (or any fields dominating the energy density of the universe) depends on the field value of the spectator field, and hence the timing of the reheating is modulated.
Spectator field models are attractive since the inflaton field itself is no more responsible for the cosmic perturbation, and hence constraints on inflation models are relaxed.
For example, the chaotic inflation~\cite{Linde:1983gd} is one of the most attractive scenarios as it is free from the initial condition problem~\cite{Linde:2005ht}.
The chaotic inflation, however, generically predicts a large tensor fraction and the simplest model is disfavored by the observations of the CMB~\cite{Ade:2015lrj}.
In spectator field models, this constraint is evaded

In this paper, we discuss the influence of a Hubble induced mass after inflation on the dynamics of the spectator field.
We focus on the curvaton and the modulated reheating scenario.
A Hubble induced mass of a field is given by the coupling of the field to the potential or kinetic energy of the inflaton.
As the potential and the kinetic energy terms are neutral under any symmetry, the Hubble induced mass is in general unsuppressed,
unless the field has an approximate shift symmetry.
Actually, the Hubble induced mass is a generic feature in supersymmetric models~\cite{Ovrut:1983my,Holman:1984yj,Goncharov:1983mw,Coughlan:1984yk}.
The Hubble induced mass of the spectator field during inflation, namely the coupling to the potential energy of the inflaton, must be suppressed to explain the almost scale invariant spectrum of the curvature perturbation.
However, the coupling of the spectator field to the kinetic energy of the inflaton is expected to be unsuppressed.
Such coupling generates a sizable Hubble induced mass after inflation.

In the curvaton scenario, the magnitude of the cosmic perturbation strongly depends on the evolutions of the inflaton and the spectator field (curvaton) after inflation.
The Hubble induced mass of the curvaton after inflation can enhance the field value of the curvaton, and the curvaton field can more easily dominate the energy density of the universe. As a result, constraints on inflation models and curvaton models, e.g.~those on the inflation scale and the reheating temperature, are considerably relaxed. This was pointed out in Refs.~\cite{McDonald:2003xq,McDonald:2004by} in the context of the leptogenesis by the sneutrino curvaton.
Here we extend the discussion to generic curvaton models, and comment on an implication to the scenarios of the baryogenesis.
In the modulated reheating scenario, as the spectator field moves during the reheating, the dependence of the timing of the reheating on the spectator field value is modified. We find that the magnitudes of the curvature perturbation and the non-gaussianity are affected.
We also propose a new model of the modulated reheating 
which is enabled by the time evolution of the spectator field due to the Hubble induced mass.
The model can realize a wide range of the local non-gaussianity parameter.

The organization of this paper is as follows.
In Sec.~\ref{sec:curvaton},
we discuss the influence of the Hubble induced mass on the curvaton scenario.
We also comment on the implications of our findings to the scenarios of baryogenesis.
In Sec.~\ref{sec:modulated}, we
investigate the modulated reheating scenario and propose a new model.
Sec.~\ref{sec:summary} is devoted to summary and discussion.

\section{Curvaton dynamics with Hubble induced mass}
\label{sec:curvaton}

In this section, we discuss the effect of the Hubble induced mass after inflation on the dynamics of the curvaton.
We first discuss the dynamics of the curvaton ignoring the Hubble induced mass.
We then take into account the effect of the Hubble induced mass. We find that the Hubble induced mass can considerably relax the constraint on curvaton models and inflation models.

\subsection{Review of the curvaton scenario}
\label{sec:review}

Before discussing the effect of the Hubble induced mass, let us review the dynamics of a curvaton ignoring the Hubble induced mass.
To be concrete, we assume that the curvaton potential is a quadratic one,%
\footnote{See Ref.~\cite{Kawasaki:2011pd} for a derivation of the curvature perturbation for a generic potential.}
\begin{align}
V(\sigma) = \frac{1}{2}m_\sigma^2  \sigma^2,
\end{align}
where $\sigma$ is a curvaton field and $m_\sigma$ is its mass.
We assume that $m_\sigma$ is much smaller than the Hubble scale during inflation, $H_I$. Then the curvaton is expected to have a
non-negligible field value during inflation, $\sigma_i$. Furthermore, the perturbation of the field value, $\delta \sigma$, is generated around $\sigma_i$,
which is the origin of the cosmic perturbation.

After inflation, the Hubble scale decreases. As it drops below $m_\sigma$, the curvaton begins its oscillation around the origin.
Eventually, the curvaton decays and the universe becomes radiation-dominant at a time $t=t_{\rm dec}$.
In the following, we assume that the inflaton decays before the curvaton decays,
otherwise the curvaton cannot dominate the energy density of the universe.
To generate large enough cosmic perturbations from the sub-dominant curvaton, the fluctuation of the curvaton energy density must be large, which leads to too large non-gaussianity.

The curvature perturbation
in the constant energy slice is fixed after the curvaton decays, and
is given by~\cite{Lyth:2002my,Lyth:2003ip}
\begin{align}
\zeta 
=
\frac{\rho_\sigma}{4 \rho_I  + 3 \rho_\sigma } \left. \frac{\delta \rho_\sigma}{\rho_\sigma}\right|_{\text{decay}}
=
 \frac{\rho_\sigma}{4 \rho_I  + 3 \rho_\sigma }  \left. \frac{2 \delta \sigma }{\sigma }\right|_{\text{decay}}
=
\frac{ f_{\rm dec}}{4 - f_{\rm dec}}  \left. \left(\frac{2 \delta \sigma}{\sigma_i}\right) \right|_\text{horizon exit},
\label{eq:zeta}
\end{align}
where $\rho_\sigma$ is the energy density of the curvaton, $\rho_I$ is the energy density of the radiation originating from the inflaton,
and $f_{\rm dec} = \left. \rho_\sigma / (\rho_I + \rho_\sigma ) \right|_\text{decay}$.
Note that $\delta\sigma / \sigma$ is constant between the horizon exit and the decay of the curvaton, as the evolution of the curvaton field value is linear.
The power spectrum of the curvature perturbation is given by
\begin{align}
{\cal P}_\zeta (k) =  \left(\frac{2 f_{\rm dec}}{4 -  f_{\rm dec}}\right)^2 \left( \frac{H_k}{2\pi \sigma_i} \right)^2,
\label{eq:power_zeta}
\end{align}
where $H_k$ is the Hubble scale at the horizon exit of the scale $k$.

Let us evaluate the energy fraction of the curvaton $f_{\rm dec}$.
We assume that the curvaton starts its oscillation before the inflaton decays.
The curvaton start its oscillation when the Hubble scale drops below $H^2 \simeq m_\sigma^2 /5 \equiv H_{\rm osc}^2$~\cite{Kawasaki:2011pd}.
The energy fraction of the curvaton at that time is given by
\begin{align}
f_{\rm osc} \simeq \frac{V (\sigma_i)}{3 \mpl^2 H_{\rm osc}^2} \simeq \frac{5}{6} \frac{\sigma_i^2}{\mpl^2}.
\label{eq:fosc}
\end{align}
The ratio is conserved until the inflaton decays into radiation, and increases afterward in proportion to the scale factor of the universe.
The fraction $f_{\rm dec}$ is then given by
\begin{align}
f_{\rm dec} \simeq
{\rm min} \left(\frac{5}{6} \frac{\sigma_i^2}{\mpl^2} \frac{T_{{\rm RH},I}}{T_{{\rm RH},\sigma}},1 \right),
\label{eq:fdec}
\end{align}
where $T_{{\rm RH},I}$ and $T_{{\rm RH},\sigma}$ are the reheating temperature by the inflaton and the curvaton, respectively.

The successful curvaton scenario requires low enough $T_{{\rm RH},\sigma}$ and/or large enough inflation scale.
For $f_{\rm dec} < 1$, by eliminating $\sigma_i$ from Eqs.~(\ref{eq:power_zeta}) and (\ref{eq:fdec}), we obtain
\begin{align}
T_{{\rm RH},\sigma} =& T_{{\rm RH},I} \times \frac{f_{\rm dec}}{(4 - f_{\rm dec})^2 } \frac{5 H_k^2}{ 6\pi^2 \mpl^2}\frac{1}{{\cal P}_\zeta(k)} \leq T_{{\rm RH},I} \times \frac{1}{9} \frac{5 H_k^2}{ 6\pi^2 \mpl^2}\frac{1}{{\cal P}_\zeta(k)} \nonumber  \\
=& 7\times 10^2~{\rm GeV} \times \frac{T_{{\rm RH},I}}{10^{9}~{\rm GeV}} \left( \frac{H_{k=0.002{\rm Mpc}^{-1}}}{10^{12}~{\rm GeV}} \right)^2,
\label{eq:temp_upp}
\end{align}
where the inequality is saturated for $f_{\rm dec}\rightarrow1$. For $f_{\rm dec}=1$, we obtain the similar bound on $T_{{\rm RH},\sigma}$.
The upper bound can be understood as the condition under which the curvaton dominates the energy density of the universe.

As we will see,
the upper bound on the reheating temperature restricts the possibility of baryogenesis.

\subsection{Hubble induced mass after inflation}

The Hubble induced mass of the curvaton during inflation,
\begin{align}
V(\sigma) = \frac{c'}{\mpl^2} V(\phi)  \frac{1}{2}\sigma^2,
\end{align}
where $\phi$ is an inflaton field and $V(\phi)$ is its potential,
should be suppressed ($|c'|\lesssim 10^{-2}$) to explain the almost scale invariant power spectrum of the curvature perturbation.
A negative $c'$ can explain the red-tilt spectrum without invoking a large first slow-roll parameter nor the hiltop-type curvaton potential (see e.g.~Refs.~\cite{McDonald:2003xq,Enqvist:2013paa,Fujita:2014iaa}.)

The Hubble induced mass after inflation, on the other hand, is not necessarily suppressed.
Even if the coupling between the curvaton and the inflaton potential is suppressed, 
the coupling with the kinetic term of the inflaton is generically un-suppressed,%
\footnote{
In multi-field inflaton models such as hybrid inflation, the potential and the kinetic energy after inflation can be dominated by a field other than the inflaton. The coupling with those energy also induce un-suppressed Hubble induced masses.
If the curvaton has an approximate shift symmetry, the Hubble induced mass after inflation is also suppressed.}
\begin{align}
- {\cal L} = \frac{c}{\mpl^2 } \partial \phi \partial \phi \times \frac{1}{2} \sigma^2,
\label{eq:kinetic}
\end{align}
with $c=O(1)$.
We note that in single field small inflation models in supergravity, the value of $c$ is determined, $c=-1/2$. (See Appendix~\ref{sec:sugra}.)

During inflation, the kinetic energy of the inflaton is negligible and hence the interaction in Eq.(\ref{eq:kinetic}) does not generate a sizable Hubble induced mass. After inflation, the inflaton begins its oscillation. Then the curvaton effectively obtains its mass,
\begin{align}
V_{\rm eff}(\sigma) = \frac{c}{\mpl^2} \vev{\partial \phi \partial \phi}_{\rm ave} \frac{1}{2}\sigma^2 = \frac{3}{2} c H^2 \sigma^2,
\label{effective mass}
\end{align}
where $\vev{\cdots}_{\rm ave}$ denotes the time-average, and we have used the relation $\vev{\partial \phi \partial \phi /2}_{\rm ave} = \rho_I/2 =3 H^2 \mpl^2/2 $.

\begin{figure}[t]
\centering
  \includegraphics[clip,width=.48\textwidth]{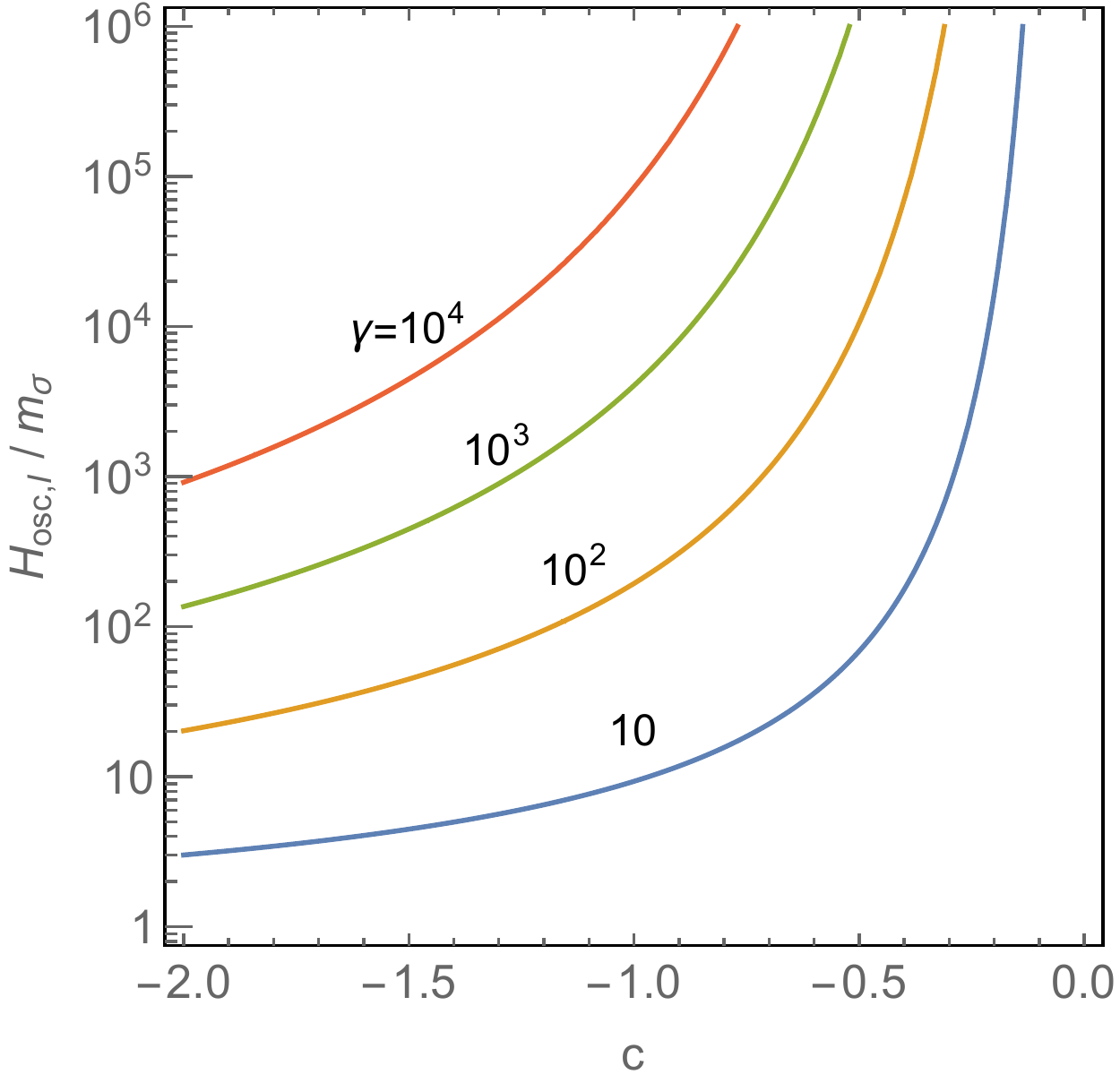}
\caption{The value of the enhancement factor $\gamma$ for given $c$ and $H_{{\rm osc},I} / m_\sigma$.}
\label{fig:gamma}
\end{figure}

The Hubble induced mass affects the dynamics of the curvaton~\cite{McDonald:2003xq,McDonald:2004by}.
For $c>0(<0)$, the curvaton field value decreases (increases) during the oscillation phase of the inflaton until the mass of the curvaton, $m_\sigma$, becomes comparable to the Hubble scale.
The equation of motion of the curvaton is given by
\begin{align}
\frac{{\rm d}^2}{{\rm d}t^2}\sigma + \frac{2}{t} \frac{{\rm d}}{{\rm d}t}\sigma + \frac{4c}{3t^2} \sigma =0.
\end{align}
The solution to this equation is
\begin{align}
\sigma (t) = \sigma_i \left( \frac{t}{t_{{\rm osc},I}}\right)^\alpha,~~\alpha = \frac{\sqrt{1 - \frac{16}{3}c}-1}{2} (> - \frac{1}{2}),
\label{sigma solution}
\end{align}
where $t_{{\rm osc},I}$ is the time at which the inflaton begins its oscillation.
Here we assume that $c < 3/16$. For larger $c$, the curvaton begins its oscillation.
As a result, the field value of the curvaton when it start its oscillation is different from $\sigma_i$,
\begin{align}
\sigma_{\rm osc} \simeq \sigma_i \times \left(\frac{H_{{\rm osc},I}}{m_\sigma / \sqrt{5}}\right)^\alpha \equiv \gamma \times \sigma_i,
\end{align}
where $H_{{\rm osc},I}$ is the Hubble scale when the inflaton begins its oscillation. Typically, it is of the same order as the Hubble scale at the end of inflation.
In Fig.~\ref{fig:gamma}, we show the value of $\gamma$ as a function of $c$ and $H_{{\rm osc},I} / m_\sigma$.
The factor $\gamma$ can be considerable.

Due to the change of the field value, discussion in Sec.~\ref{sec:review} is modified.
Eqs~(\ref{eq:zeta}) and (\ref{eq:power_zeta}) are unchanged, but $\sigma_i$ is Eqs.~(\ref{eq:fosc}) and (\ref{eq:fdec}) should be replaced with $\gamma \times \sigma_i$.
As a result, the upper bound on $T_{{\rm RH},\sigma}$ is considerably relaxed if $c<0$,
\begin{align}
T_{{\rm RH},\sigma} < 7\times 10^6~{\rm GeV} \times \frac{T_{{\rm RH},I}}{10^{9}~{\rm GeV}} \left( \frac{H_{k=0.002{\rm Mpc}^{-1}}}{10^{12}~{\rm GeV}} \right)^2 \left( \frac{\gamma}{100} \right)^2.
\label{eq:TRH_hubble}
\end{align}

\subsection{Implication to baryogenesis}

Here we discuss the implications of our findings in the previous section to scenarios of baryogenesis.
In the curvaton scenario, the baryon or lepton asymmetry should be produced after the curvaton dominates the energy density of the universe~\cite{Lyth:2002my,Lyth:2003ip,Ikegami:2004ve}.
Otherwise, large baryon isocurvature perturbations are produced, which is excluded by the constraint from the observations of the CMB.%
\footnote{The constraint from the CMB can be evaded if the baryon and the dark matter isocurvature perturbation cancel with each other~\cite{Gordon:2002gv,Harigaya:2012up,Harigaya:2014bsa}.}
Here we focus on
the thermal leptogenesis~\cite{Fukugita:1986hr},
the non-thermal leptogenesis from the decay of right-handed sneutrino curvaton~\cite{Murayama:1993em,Moroi:2002vx,Postma:2002et}
and the Affleck-Dine (AD) baryogenesis~\cite{Affleck:1984fy,Dine:1995kz}.

\subsubsection{Thermal leptogenesis}
\begin{figure}[t]
\centering
  \includegraphics[clip,width=.48\textwidth]{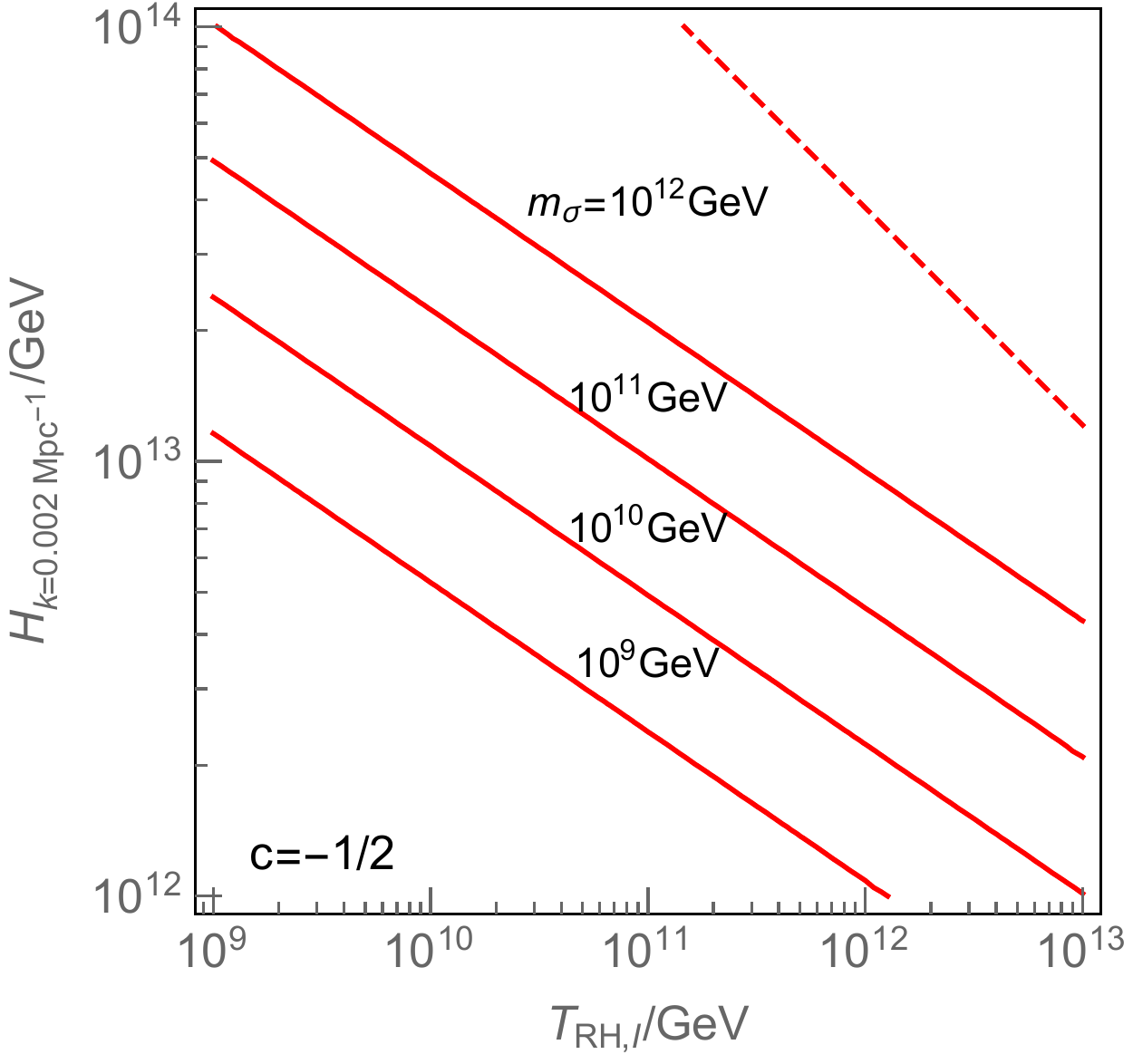}
    \includegraphics[clip,width=.48\textwidth]{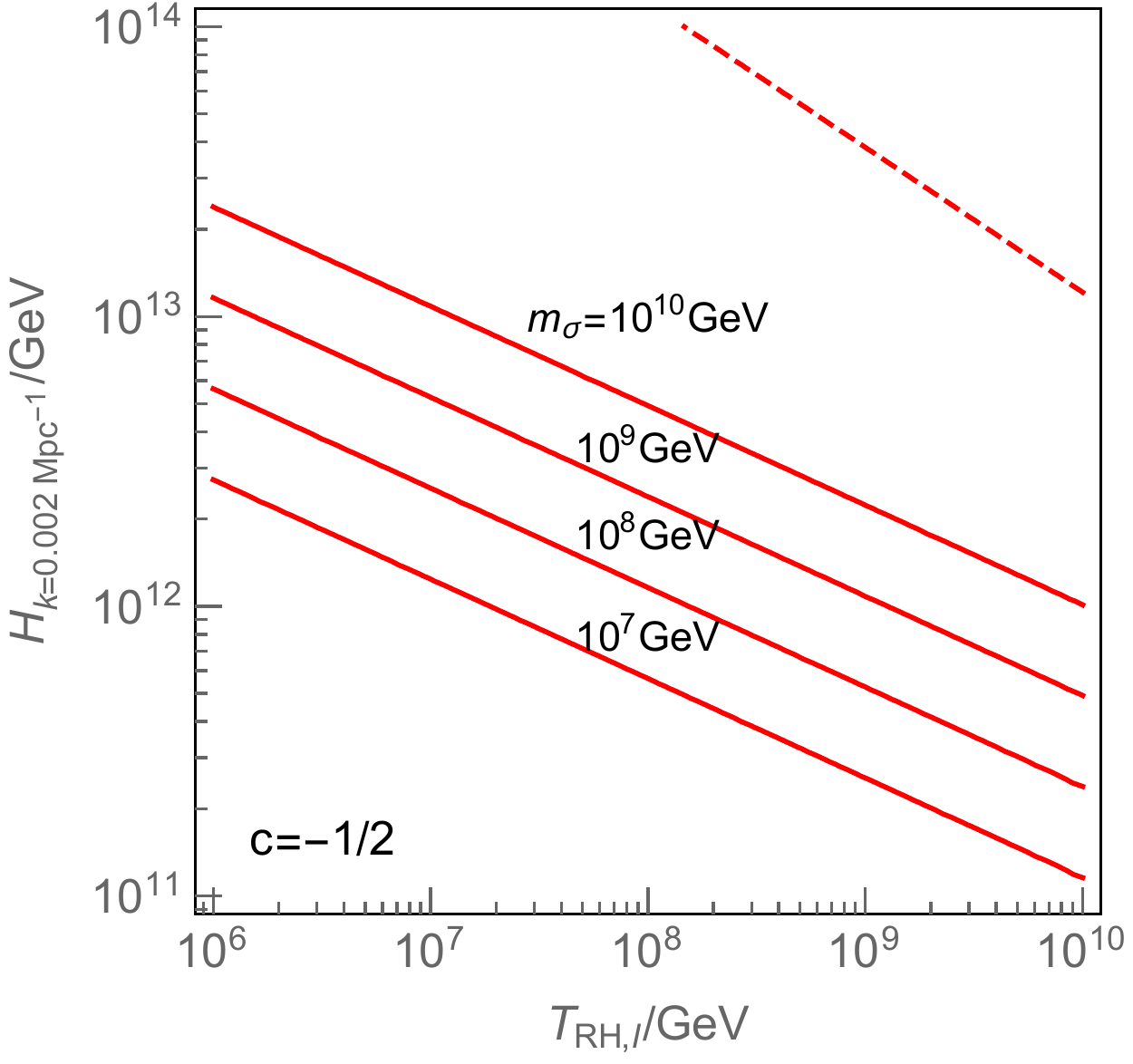}
\caption{The constraint on the reheating temperature of the inflaton, $T_{{\rm RH},I}$, and the Hubble scale during inflation at the pivot scale, $H_{k=0.002{\rm Mpc}^{-1}}$.
Below the solid lines of the left panel, the upper bound on $T_{{\rm RH},\sigma}$ is smaller than $10^9$ GeV, and the thermal leptogenesis cannot produce sufficient amount of the baryon asymmetry. Here we take $c=-1/2$. For the dashed line, we take $c=0$.
The light panel shows the similar constraint for the successful leptogenesis by the decay of the sneutrino curvaton, which requires that $T_{{\rm RH},\sigma}>10^6$ GeV.
}
\label{fig:leptogenesis}
\end{figure}

The thermal leptogenesis requires that $T_{{\rm RH},\sigma} \gtrsim 10^9$ GeV~\cite{Buchmuller:2004nz}.
In Fig.~\ref{fig:leptogenesis}, we show the constraint on the reheating temperature of the inflaton, $T_{{\rm RH},I}$, and the Hubble scale during inflation at the pivot scale $H_{k=0.002{\rm Mpc}^{-1}}$.
Below the solid lines of the left panel, the upper bound on $T_{{\rm RH},\sigma}$ in Eq.~(\ref{eq:TRH_hubble}) is lower than $10^9$ GeV. Here, we take $c=-1/2$ and assume that $H_{{\rm osc},I}\simeq H_{k=0.002{\rm Mpc}^{-1}}$. For the dashed line, we take $c=0$. It can be seen that the effect of the Hubble induced mass extends the parameter region consistent with the thermal leptogenesis.

\subsubsection{Sneutrino curvaton}
Next, we consider the leptogenesis by the decay of the right-handed sneutrino curvaton.
in order to produce a sufficient amount of the lepton asymmetry, it is required that $T_{{\rm RH},\sigma}\gtrsim 10^6$ GeV~\cite{Hamaguchi:2001gw}.
Below the solid lines of the right panel of Fig.~\ref{fig:leptogenesis}, the upper bound on $T_{{\rm RH},\sigma}$ in Eq.~(\ref{eq:TRH_hubble}) is lower than $10^6$ GeV. Here, we take $c=-1/2$. For the dashed line, we take $c=0$. It can be seen that the effect of the Hubble induced mass extends the parameter region consistent with the leptogenesis by the sneutrino curvaton~\cite{McDonald:2003xq,McDonald:2004by}.

\subsubsection{The Affleck-Dine baryogenesis}
Finally, we consider the AD baryogenesis scenario by flat directions in the minimal supersymmetric standard model.
As the efficiency of the AD baryogenesis depends on the detailed property of the flat direction as well as the mediation scheme of the supersymmetry breaking, we comment only on the constraint applicable to generic models.
The AD field should start its oscillation after the curvaton dominates the energy density of the universe.
The Hubble scale when the curvaton dominates the energy density of the universe is given by
\begin{align}
H_{\rm dom} = \frac{25 \gamma^4 \sigma_i^4}{36 \mpl^4} \sqrt{\frac{\pi^2}{90} g_*} \frac{T_{{\rm RH},I}^2}{\mpl} = \frac{25 \gamma^4}{2916 \pi^4 {\cal P}_\zeta(k)^2} \sqrt{\frac{\pi^2}{90} g_*} \left( \frac{H_k}{\mpl} \right)^4 \frac{T_{{\rm RH},I}^2}{\mpl} \nonumber \\
= 100~{\rm GeV} \times  \left(\frac{T_{{\rm RH},I}}{10^{10}~{\rm GeV}}\right)^2 \left( \frac{H_{k=0.002{\rm Mpc}^{-1}}}{10^{13}~{\rm GeV}} \right)^4\left( \frac{\gamma}{100} \right)^4.
\end{align}
Since the AD field starts its oscillation when the Hubble scale drops below the mass of the AD field,
the mass of the AD field must be smaller than $H_{\rm dom}$.
For example, in gravity mediated models, where the mass of the AD filed is as large as the masses of superparticles around the origin, $O(1)$ TeV, 
the reheating temperature of the inflaton as well as the Hubble scale must be large.
The enhancement of the curvaton field value by the negative Hubble induced mass relaxes those constraints.

\section{Modulated reheating with Hubble induced mass}
\label{sec:modulated}

In this section, we discuss the effect of the Hubble induced mass after inflation on a spectator scalar field whose field value determines 
the timing of the inflaton decay.  We briefly review the case without the Hubble
induced mass and then take into account its effect.
We derive an analytic result under the sudden decay approximation as well as a result without the approximation by numerical calculations.
It is found that the prediction for the curvature perturbation and the non-gaussianity can be  considerably changed due to the Hubble induced mass.
We also propose a new model of the modulated reheating utilizing the Hubble induced mass.

\subsection{Review of the modulated reheating}

In the modulated reheating scenario, the curvature perturbation originates from the fluctuation of a spectator field $\sigma$.
Contrary to curvaton models, the energy density of the spectator field may be always subdominant in this scenario, which we assume in this paper.
If the decay rate of the inflaton $\Gamma_\phi$ depends on the field value of $\sigma$, the transition time from the inflaton oscillating era to the radiation dominated era is perturbed by the fluctuation of $\sigma$.
Under the sudden decay approximation which assumes that the inflaton decays instantaneously at $H=\Gamma_\phi$, the e-folding number between the onset of the inflaton oscillation ($H=H_{{\rm osc},I}$) and a sufficiently later time after its decay ($H=H_f$) is given by
\begin{equation}
\Delta N
= \frac{2}{3}\ln \left(\frac{H_{{\rm osc},I}}{\Gamma_\phi}\right)
+\frac{1}{2}\ln \left(\frac{\Gamma_\phi}{H_f}\right),
\label{Delta N}
\end{equation}
Based on the $\delta N$ formalism~\cite{Sasaki:1995aw,Wands:2000dp,Lyth:2004gb}, one can show that the contribution to the curvature perturbation and the local non-gaussianity from the fluctuation of $\sigma$ are
\begin{equation}
\zeta =(\partial_\s \D N) \d\s = -\frac{1}{6} \left(\partial_\s \ln \G_\phi\right)\d\s,
\qquad
f_{\rm NL} =\frac{5}{6}\frac{\partial_\sigma^2 \Delta N}{(\partial_\s \D N)^2} = -5 \frac{\partial_\s^2 \ln\G_\phi}{(\partial_\s \ln\G_\phi)^2}, 
\end{equation}
where $\partial_\s$ denotes 
the derivative with respect to $\s$.
Here we have assumed that $\sigma$ acquires a gaussian fluctuation $\d\s$ during inflation and the field value of $\s$ does not significantly evolve after inflation. 
For instance, if $\Gamma_\phi$ is a power-law function of $\sigma$,  
these observables are computed as
\begin{equation}
\Gamma_\phi = g \sigma^n
\qquad\Longrightarrow\qquad
\zeta = -\frac{n}{6} \frac{\delta\s}{\s},\quad
f_{\rm NL} = \frac{5}{n}.
\label{conventional results}
\end{equation}
This is a typical result of the conventional modulated reheating scenario.
As we saw in the previous section, however, the Hubble induced mass can cause a non-trivial time evolution of $\s$.

\subsection{Field-dependent coupling}

Now we consider the case where $\sigma$ has a Hubble induced mass, Eq.~\eqref{effective mass}, and $\Gamma_\phi=g \s^n$. Since it evolves as $\s\propto t^\a$ (see Eq.~\eqref{sigma solution}), $\Gamma_\phi$ equals $H$ when $H$ goes down to
\begin{equation}
H_d = \left(\frac{\Gamma_{{\rm osc},I}}{H_{{\rm osc},I}}\right)^{\frac{1}{n\a+1}}H_{{\rm osc},I},
\end{equation}
where $\Gamma_{{\rm osc},I}=g \s^n(t_{{\rm osc},I})$.
Replacing $\Gamma_\phi$ in Eq.~\eqref{Delta N} by this $H_d$,
we obtain 
\begin{align}
\zeta= -\frac{1}{6}\frac{n}{n\alpha+1} \frac{\delta\sigma}{\sigma},\qquad
f_{\rm NL}
=5 \frac{n\alpha+1}{n}.
\end{align}
Compared with the conventional result in Eq.~\eqref{conventional results},
$\zeta$ and $f_{\rm NL}$ are divided and multiplied by $n\a+1$, respectively.
Therefore the effect of the Hubble induced mass cannot be ignored unless
$|n\a|\ll1$.
Note that $\zeta$ formally  diverges if $\a=-1/n$. It is because $\G_\phi (\propto t^{n\alpha})$ and $H (\propto t^{-1})$ decrease at the same rate and the inflaton never decays in that case. 

In order to obtain the predictions for $\zeta$ and $f_{\rm NL}$ without the sudden decay approximation, we numerically solve the following equations:
\begin{align}
&\partial_N^2 \sigma +\left(3+\partial_N \ln H\right)\partial_N \sigma+\frac{c\rho_\phi+\tilde{c} \rho_r}{\Mpl^2 H^2} \sigma=0,
\\
&\partial_N \rho_\phi +3\rho_\phi = -\frac{\Gamma_\phi}{H}\rho_\phi,
\qquad
\partial_N \rho_r +4\rho_r = \frac{\Gamma_\phi}{H}\rho_\phi,
\end{align}
where $\partial_N$ denotes the derivative with respect to e-folding number,
$\rho_\phi$ and $\rho_r$ are the energy density of the inflaton and radiation, respectively, and $\tilde c$ is the coefficient of the coupling between $\s$ and the radiation energy density, $\mathcal{L}=-\tilde{c} \rho_r \s^2/2\Mpl^2$.
For Planck-suppressed couplings between $\sigma$ and particles in the thermal bath,
$\tilde c$ is typically $\mathcal{O}(10^{-2}-10^{-3})$~\cite{Kawasaki:2011zi,Kawasaki:2012qm,Kawasaki:2012rs}.
We use $\tilde{c}=10^{-2}$ hereafter. It is assumed that the energy density of $\s$ is negligible and $3\Mpl^2 H^2 =\rho_\phi+\rho_r$. The initial condition is $\rho_r(t_{{\rm osc},I}) =0$ and $\partial_N \s(t_{{\rm osc},I}) = \frac{3}{2} \a \s_{{\rm osc},I}$.\footnote{This initial condition of $\partial_N\s$ ensures that Eq.~\eqref{sigma solution} is correct and another linearly independent, more rapidly decreasing solution vanishes. It is straightforward to extend the analysis to a more general initial condition.}

\begin{figure}[t]
\centering
  \includegraphics[clip,width=.48\textwidth]{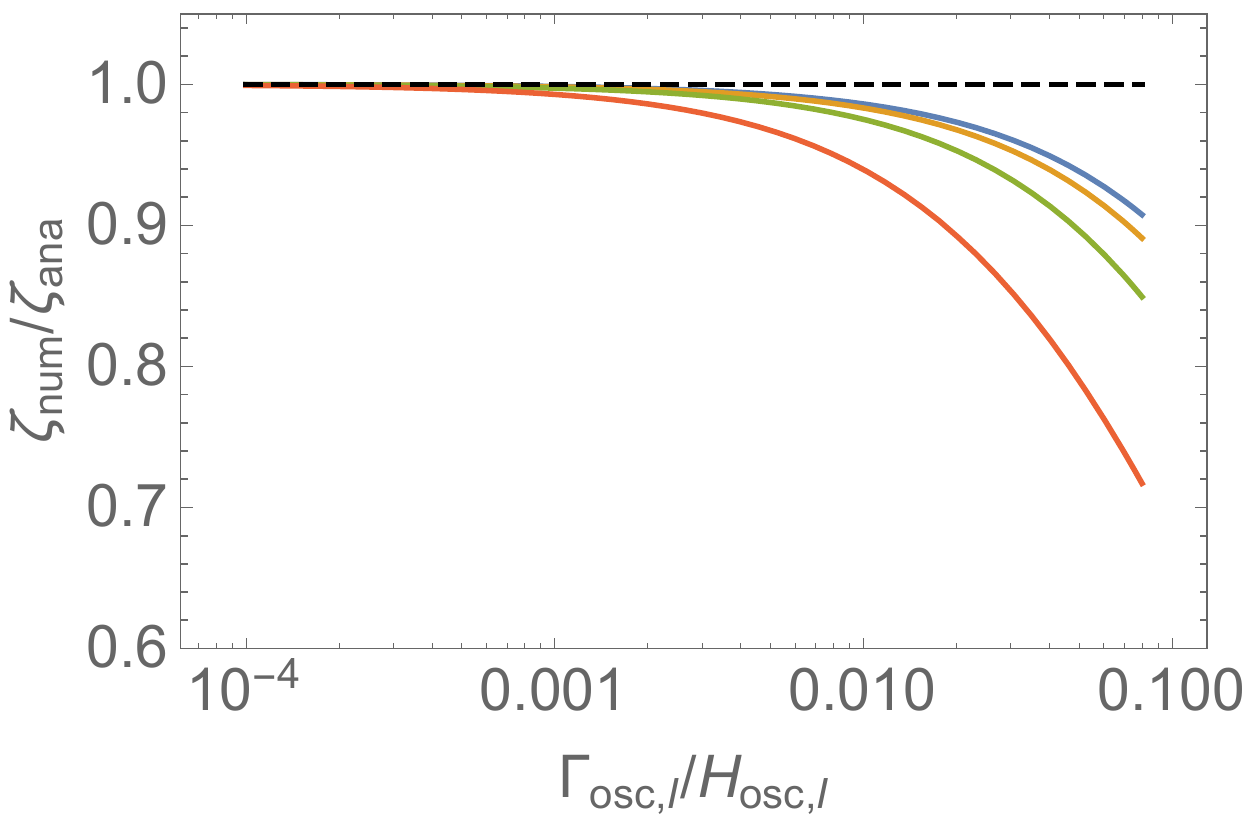}
    \includegraphics[clip,width=.48\textwidth]{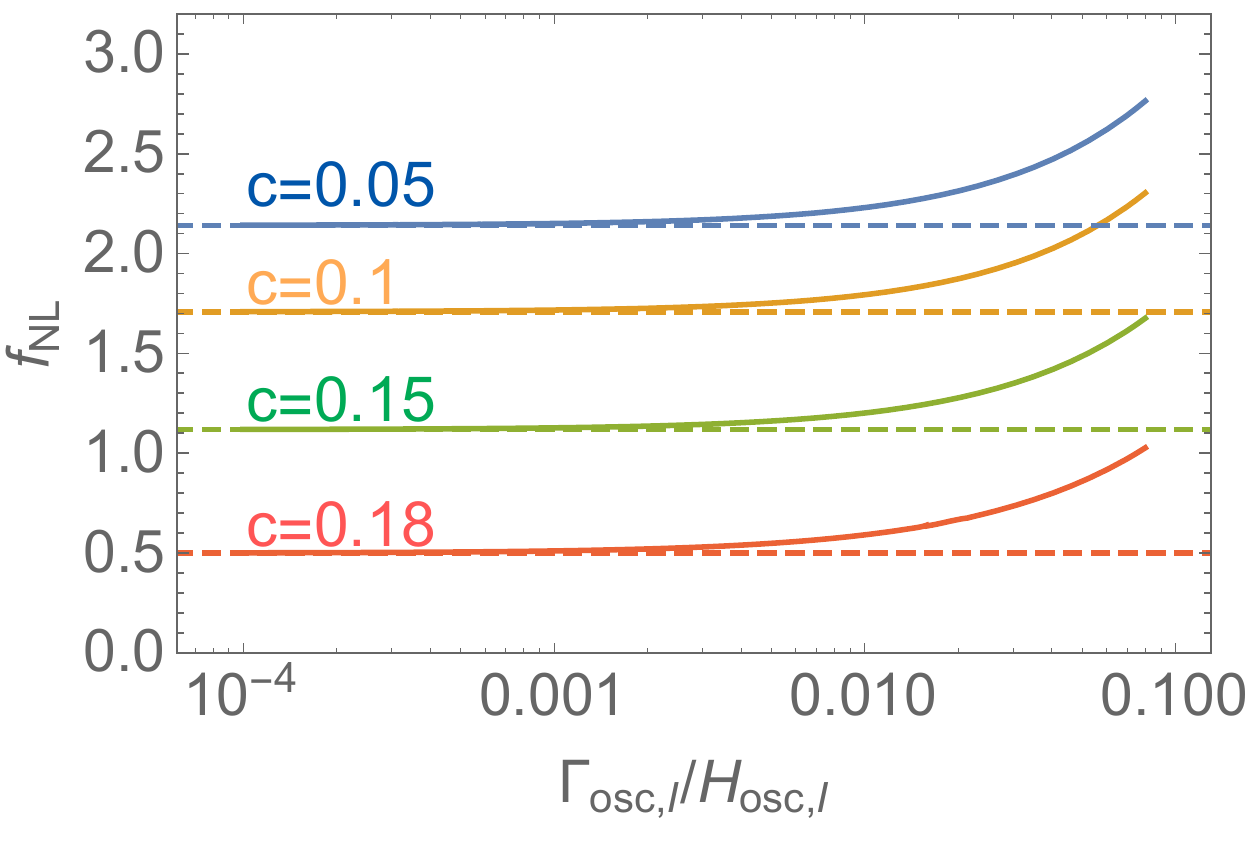}
\caption{
The comparison between the analytic expressions derived under the sudden decay approximation and the numerical results
in the modulated reheating scenario with the Hubble induced mass. 
We fix $n=2$ and $c=0.05$ (blue), $0.1$ (yellow), $0.15$ (green) and $0.18$ (red). The solid lines are the numerical results while the dashed lines are the analytic ones. The left and right panel show the ratio of $\zeta$ and the local non-gaussianity $f_{\rm NL}$, respectively.
The sudden decay approximation becomes less accurate for larger decay rate and larger Hubble induced mass.
}
\label{fig:decayrate}
\end{figure}

Running a numerical calculation with changing an initial value of $\G_\phi$ (namely $\G_{{\rm osc},I}$), one obtains $\D N$ as a function of $\Gamma_{{\rm osc},I}$. Then $\zeta$ and $f_{\rm NL}$ can be computed.
To compare the numerical results and the analytic expressions,
we plot the following quantities in Fig.~\ref{fig:decayrate}.
\begin{align}
\frac{\zeta_{\rm num}}{\zeta_{\rm ana}} = -6(n\alpha+1) \G_{{\rm osc},I}\partial_\Gamma \Delta N,
\qquad
f_{\rm NL}= \frac{5}{6}\frac{\partial_\G^2\Delta N+\frac{n-1}{n \G_{{\rm osc},I}}\partial_\G \Delta N}{(\partial_\G \Delta N)^2},
\end{align}
where $\partial_\G$ is the derivative with respect to $\G_{{\rm osc},I}$.

As one can see in Fig.~\ref{fig:decayrate}, the analytic expressions derived under the sudden decay approximation become less accurate as the Hubble induced mass or the initial value of $\G_\phi$ increases. One expects that the approximation gets worse, as  $c$ increases and $\a$ approaches $-1/n$, because $\Gamma_{\phi}$ and $H$ are comparable for a longer time.
In addition, a large $\G_{{\rm osc},I}$ also invalidates the approximation; if $\G_\phi/H$ is not negligible at the beginning, the radiation component is significant most of the time before $H=\G_\phi$, and the deviation from the inflaton oscillation era ($a\propto t^{2/3}$) becomes relevant.
For negative $c$ (positive $\a$), however, 
the analytic and numerical results are in good agreement. It is because $\G_\phi$ increases in this case and the sudden decay approximation is more accurate.

\subsection{Field-dependent mass of inflaton decay products}

In the conventional modulated reheating scenario, the time evolution of $\sigma$ is ignored and the fluctuation of $\s$ is converted into $\zeta$ through the inflaton decay rate depending on $\s$.
Nevertheless, the time evolution of $\s$ can be significant in the case with the Hubble induced mass and we find the following new mechanism of generating $\zeta$.%
\footnote{A conceptually similar mechanism to produce curvature perturbations through the evaporation of primordial black holes is discussed in Refs.~\cite{Fujita:2013bka,Fujita:2014hha}.}

If the inflaton decays into a particle $\psi$ and its mass $m_\psi$ depends on $\s$,  the time evolution of $\s$ given in Eq.~\eqref{sigma solution} can open the on-shell decay channel of the inflaton which is initially closed. 
In this paper, we consider that the inflaton decays into two $\psi$ particles, $\phi\rightarrow\psi \psi$, and the mass of $\psi$
depends on $\s$ as
\begin{align}
m_\psi = y\sigma^n.
\end{align}
Provided that  $2m_{\p}=2 y \s^n > m_\phi$ at the end of inflation and $\s$ decreases after inflation with $\alpha<0$, the  decay of the inflaton is kinematically prohibited at the beginning of the inflaton oscillating era,%
\footnote{In the following we assume that $\psi$ decays into radiation immediately after the inflaton decays into $\psi$s.
Then the decay of the inflaton through off-shell $\psi$ is possible even for $2m_{\p} > m_\phi$.
The decay rate is suppressed in comparison with $\Gamma_\phi$ by the multi body factor as well as the coupling between $\psi$ and radiation, and can be neglected.
}
while it is allowed once $\s$ decreases to the critical value, $\s_d\equiv(m_\phi/2y)^{1/n}$. In this case, the inflaton decay rate is written as
\begin{equation}
\Gamma_\phi = \Gamma \, \Theta\left( \sigma_d-\sigma(t)\right)\sqrt{1-\frac{4m_\psi^2}{m_\phi^2}},
\end{equation}
where the Heaviside function $\Theta$ and the square-root factor represent the kinematic prohibition and the kinematic suppression effect, respectively.
Here we assume that the martix element of the decay process is not suppressed by the momentum of the final state.
Inclusion of such possible factor is straightforward.
Assuming that other decay channels are suppressed, the Hubble parameter at the onset of the inflaton decay can be obtained as
\begin{equation}
H_d = H_{{\rm osc},I}\left(\frac{\sigma_d}{\sigma_{{\rm osc},I}}\right)^{-1/\a}.
\end{equation}
Substituting it into $\D N$, one finds
\begin{align}
\zeta = -\frac{1}{6\alpha}\frac{\delta\sigma}{\s},
\qquad
f_{\rm NL}= 5\alpha.
\end{align}
In the limit $\a \to 0$, $\zeta$ formally diverges. This is because $\s\propto t^{\a}$ stops evolving and this mechanism cease to work.
It should be noted that the sudden decay approximation is justified if $\Gamma/H_d\gg 1$.

\begin{figure}[t]
\centering
  \includegraphics[clip,width=.48\textwidth]{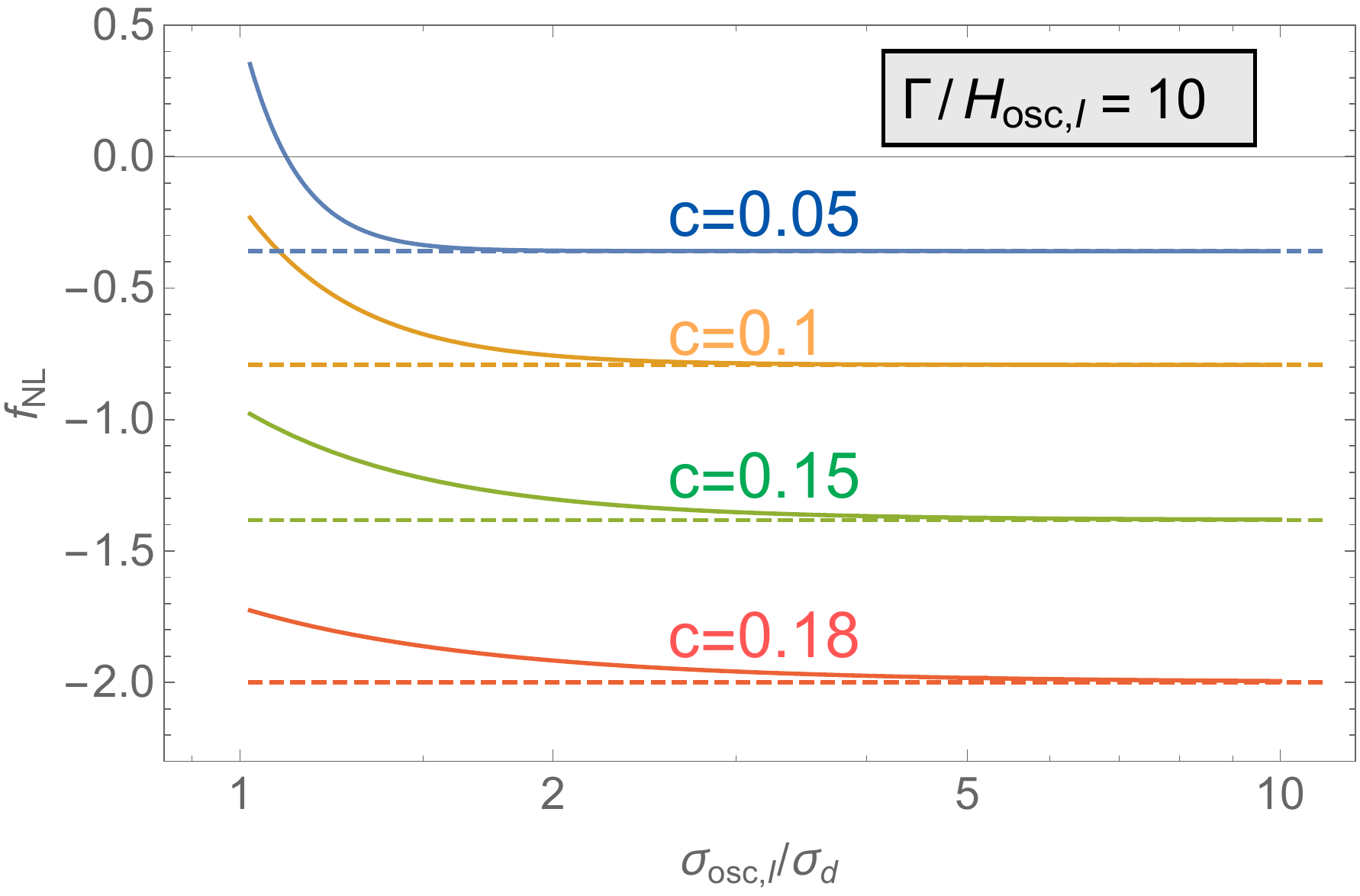}
    \includegraphics[clip,width=.48\textwidth]{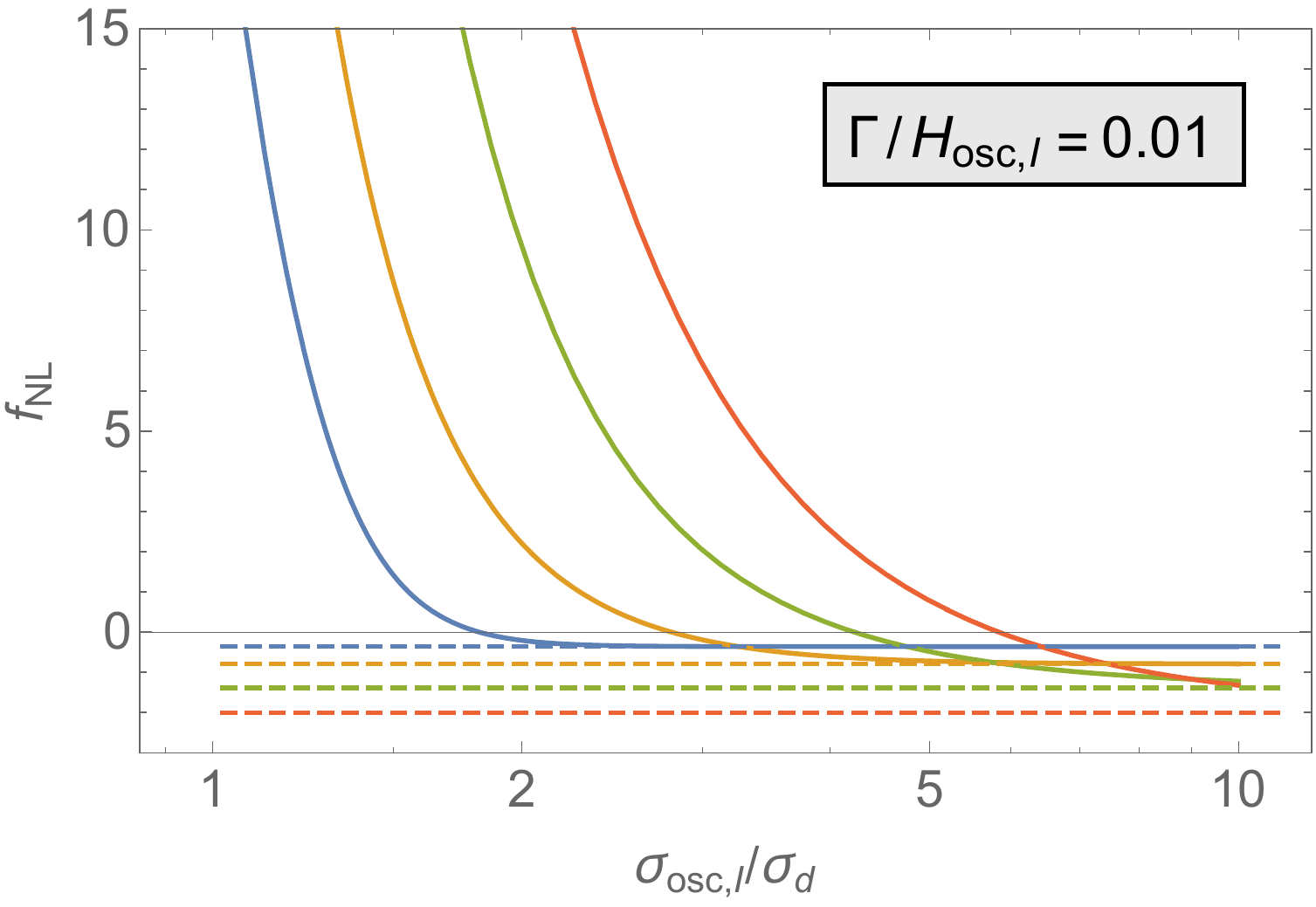}
\caption{
The comparison between the analytic expressions derived under the sudden decay approximation and the numerical results
in the new type of the modulated reheating scenario with the Hubble induced mass. 
We fix $n=2$ and $c=0.05$ (blue), $0.1$ (yellow), $0.15$ (green) and $0.18$ (red). The solid lines are the numerical results while the dashed lines are the analytic ones. The left and right panel show the local non-gaussianity $f_{\rm NL}$ in the case with $\G/H_{{\rm osc},I}=10$ and $10^{-2}$, respectively.
The sudden decay approximation is valid if $\G/H_d\gg 1$. }
\label{fig:mass}
\end{figure}

In Fig.~\ref{fig:mass}, we show the comparison of the analytic result of $f_{\rm NL}$ with the numerical one. If $\s_{{\rm osc},I}/\s_d$ is large, while $\s$ goes down to $\s_d$, $H$ significantly decreases and $\G/H_d$ becomes large. On the other hand, if $\G/H_{{\rm osc},I}$ is small and $\s_{{\rm osc},I}/\s_d$ is close to unity, $\G/H_d$ cannot be large enough to justify the sudden decay approximation.%
\footnote{In this case, the inflaton decay is not effective even when $\sigma$ reaches $\sigma_d$ and the on-shell decay channel opens. Since the dependence of $\D N$ on $\sigma$ becomes weak, $\zeta$ is suppressed and $f_{\rm NL}$ is boosted as seen in Fig.~\ref{fig:mass}.}
Contrary to the analytic prediction, $0>f_{\rm NL}^{\rm ana}=5\a>-5/2$, the numerical computation indicates that positive $f_{\rm NL}$ can be produced in such cases.
Indeed, Fig.~\ref{fig:mass} shows that an arbitrary value of the local non-gaussianity between $-5/2\lesssim f_{\rm NL} \lesssim 10$ can be realized in this new mechanism of the modulated reheating scenario, depending on the parameters.

\begin{figure}[t]
\centering
    \includegraphics[clip,width=.5\textwidth]{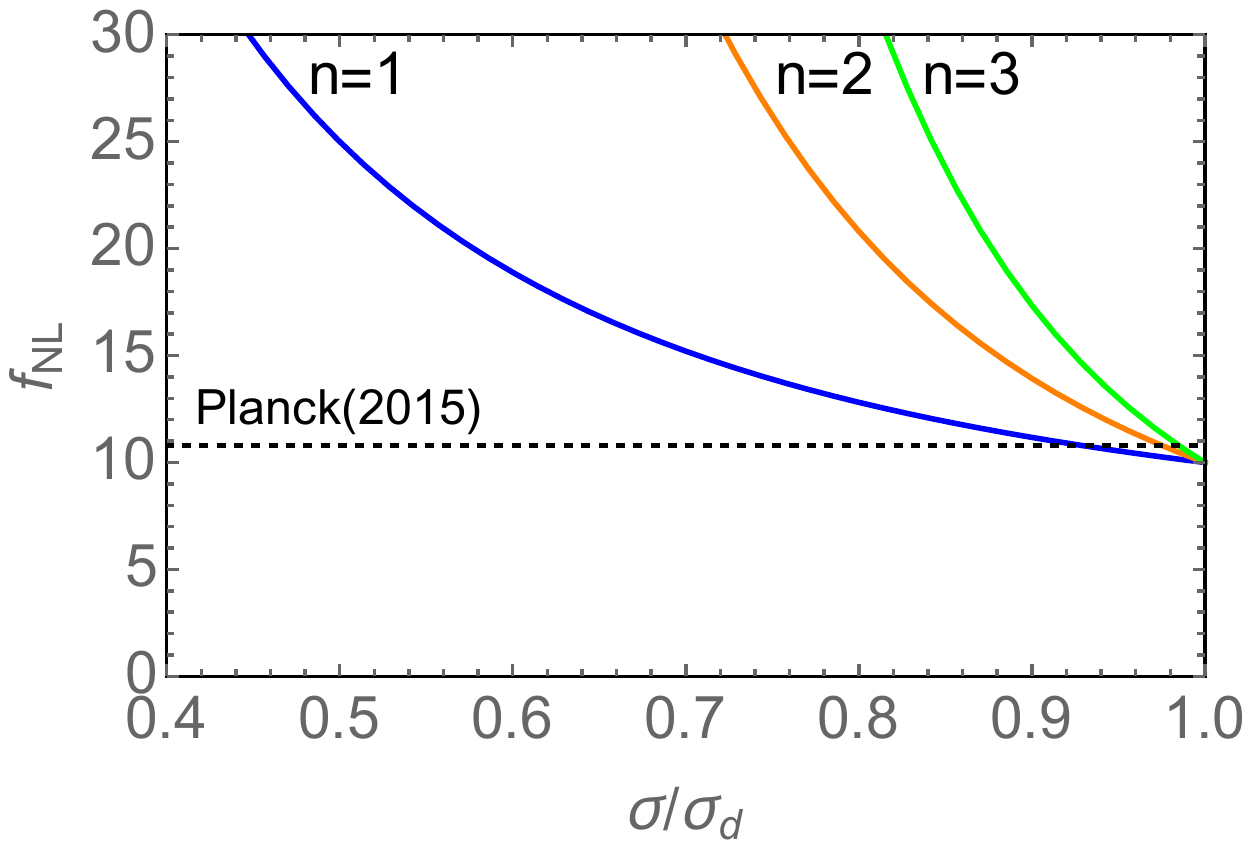}
\caption{
The prediction on the local non-gaussianity parameter $f_{\rm NL}$ as a function of $\sigma$ with neglecting the Hubble induced mass.
The dashed line shows the upper bound on $f_{\rm NL}$~\cite{Ade:2015ava}.
}
\label{fig:fnl_sigma}
\end{figure}

Let us comment on an implication of our result in the models of the modulated reheating.
It is a common feature in the standard model as well as the beyond standard model that a mass of a particle depends on a value of some field.
It would be thus natural to consider a model of modulated reheating where the mass of the daughter particle of the inflaton is modulated.
When the Hubble induced mass of the spectator field is neglected, the modulation originates only from the kinetic suppression factor,
\begin{align}
\Gamma_\phi = \Gamma \sqrt{ 1 - \frac{4 m_\psi^2}{m_{\phi}^2} },~~m_\psi = y \sigma^n.
\label{eq:decay_kinetic}
\end{align}
Then the non-gaussianity parameter $f_{\rm NL}$ is given by
\begin{align}
f_{\rm NL} = \frac{5}{n} \left( \frac{\sigma_d}{\sigma} \right)^{2n}
\left(
\left( 2n-1 \right) + \left( \frac{\sigma}{\sigma_d} \right)^{2n}
\right).
\end{align}
In Fig.~\ref{fig:fnl_sigma}, we show the prediction on $f_{\rm NL}$ as a function of $\sigma$ for $n=1,2,3$.
The upper bound on $f_{\rm NL}$, $f_{\rm NL}<10.8$~\cite{Ade:2015ava} (indicated by a dashed line), is satisfied only when $\sigma$ is accidentally close to $\sigma_d$,
which makes the modulated reheating with a modulated mass less interesting.%
\footnote{Multi-body decay of the inflaton and/or the momentum suppressed matrix element of the decay can change the kinetic factor in Eq.~(\ref{eq:decay_kinetic}) and suppress $f_{\rm NL}$ by few factor. Still, it is necessary that $\sigma$ is close to some specific value.}

As we have shown, with the Hubble induced mass, the non-gaussianity is suppressed as long as $\sigma > \sigma_d$ initially.
This shows the importance of the Hubble induced mass in models of the modulated reheating.

\section{Summary}
\label{sec:summary}
In this paper, we have discussed the effect of a Hubble induced mass after inflation on the dynamics of spectator field models of the generation of the cosmic perturbation.
Although the Hubble induced mass of the spectator field must be suppressed during inflation, it can be sizable after inflation.
Actually, in supersymmetric inflation models, it is in general unsuppressed. 

We have found that the Hubble induced mass can enhance the field value of the curvaton, which helps the curvaton to dominate the energy density of the universe.
As a result, the constraints on inflation models and curvaton models are considerably relaxed.
We have also discussed the implications of our finding to the scenarios of baryogenesis.
The possibility of the baryogenesis is extended.

For the modulated reheating scenario, we find that the magnitudes of the curvature perturbation and the non-gaussianity are affected by the Hubble induced mass.
We also propose a new model of the modulated reheating utilizing the Hubble induced mass,
which can produce an arbitrary value of the local non-gaussianity parameter.

\section*{Acknowledgements}
This work was supported in part by the Japan Society for the Promotion of Science Postdoctoral Fellowships for Research Abroad, Grant No. 27-154 (T.~F.),
by the Department of Energy, Office of Science, Office of High Energy Physics, under contract No. DE-AC02-05CH11231 (K.~H.), and by the National Science Foundation under grants PHY-1316783 and PHY-1521446 (K.~H.)

\appendix

\section{Hubble induced mass in supergravity}
\label{sec:sugra}

In this appendix, we discuss the coupling between a spectator field and the kinetic term of an inflaton in supergravity.
Let us first consider a small field inflation model with a single chiral field.
Let $\Phi$ and $N$ be the chiral multiplet whose lowest component include an inflaton $\phi$ and a spectator field $\sigma$, respectively.
Assuming that the field value of $\sigma$ is much smaller than the Planck scale,
the following Kahler potential is relevant for the Hubble induced mass of the spectator field,
\begin{align}
K = \Phi^\dag \Phi + N^\dag N + \frac{d}{\mpl^2} \Phi^\dag \Phi N^\dag N.
\end{align}
In single field inflation models, the chiral field $\Phi$ obtains a non-zero $F$ term during inflation. Thus, the Hubble induced mass of the spectator field during inflation is given by
\begin{align}
V(\phi,\sigma) = \frac{1}{\mpl^2} V (\phi) (1-d) \frac{1}{2} \sigma^2.
\end{align}
To suppress the Hubble induced mass, the constant $d$ must be close to unity.
On the other hand, the derivative couplings between $\phi$ and $\sigma$ is given by
\begin{align}
{\cal L} = d\times \left [ \frac{1}{2} \sigma^2 \frac{1}{2} \partial \phi \partial \phi + \frac{1}{2} \phi^2 \frac{1}{2} \partial \sigma \partial \sigma + \frac{1}{2} \sigma \phi \partial \sigma \partial \phi \right].
\label{eq:kinetic couplings}
\end{align}
The second and the third terms are negligible as long as $|\phi| \ll \mpl$.
Comparing Eqs.~(\ref{eq:kinetic couplings}) and (\ref{eq:kinetic}), we obtain $c \simeq -1/2$.

In a large field inflation models where the inflaton field value changes by $O(\mpl)$, higher order Kahler potential terms are also relevant for the Hubble induced mass term of the spectator field, and hence $c$ is an $O(1)$ free parameter.   
In inflation models with multiple chiral field, a chiral field which has a non-zero $F$ term during inflation is in general different from a chiral field containing an inflaton. Thus, the constant $c$ is again an $O(1)$ free paramter.

\newpage

\bibliography{spectator}

\begin{thebibliography}{47}%
\makeatletter
\providecommand \@ifxundefined [1]{%
 \@ifx{#1\undefined}
}%
\providecommand \@ifnum [1]{%
 \ifnum #1\expandafter \@firstoftwo
 \else \expandafter \@secondoftwo
 \fi
}%
\providecommand \@ifx [1]{%
 \ifx #1\expandafter \@firstoftwo
 \else \expandafter \@secondoftwo
 \fi
}%
\providecommand \natexlab [1]{#1}%
\providecommand \enquote  [1]{``#1''}%
\providecommand \bibnamefont  [1]{#1}%
\providecommand \bibfnamefont [1]{#1}%
\providecommand \citenamefont [1]{#1}%
\providecommand \href@noop [0]{\@secondoftwo}%
\providecommand \href [0]{\begingroup \@sanitize@url \@href}%
\providecommand \@href[1]{\@@startlink{#1}\@@href}%
\providecommand \@@href[1]{\endgroup#1\@@endlink}%
\providecommand \@sanitize@url [0]{\catcode `\\12\catcode `\$12\catcode
  `\&12\catcode `\#12\catcode `\^12\catcode `\_12\catcode `\%12\relax}%
\providecommand \@@startlink[1]{}%
\providecommand \@@endlink[0]{}%
\providecommand \url  [0]{\begingroup\@sanitize@url \@url }%
\providecommand \@url [1]{\endgroup\@href {#1}{\urlprefix }}%
\providecommand \urlprefix  [0]{URL }%
\providecommand \Eprint [0]{\href }%
\providecommand \doibase [0]{http://dx.doi.org/}%
\providecommand \selectlanguage [0]{\@gobble}%
\providecommand \bibinfo  [0]{\@secondoftwo}%
\providecommand \bibfield  [0]{\@secondoftwo}%
\providecommand \translation [1]{[#1]}%
\providecommand \BibitemOpen [0]{}%
\providecommand \bibitemStop [0]{}%
\providecommand \bibitemNoStop [0]{.\EOS\space}%
\providecommand \EOS [0]{\spacefactor3000\relax}%
\providecommand \BibitemShut  [1]{\csname bibitem#1\endcsname}%
\let\auto@bib@innerbib\@empty
\bibitem [{\citenamefont {Mukhanov}\ and\ \citenamefont
  {Chibisov}(1981)}]{Mukhanov:1981xt}%
  \BibitemOpen
  \bibfield  {author} {\bibinfo {author} {\bibfnamefont {V.~F.}\ \bibnamefont
  {Mukhanov}}\ and\ \bibinfo {author} {\bibfnamefont {G.~V.}\ \bibnamefont
  {Chibisov}},\ }\href@noop {} {\bibfield  {journal} {\bibinfo  {journal} {JETP
  Lett.}\ }\textbf {\bibinfo {volume} {33}},\ \bibinfo {pages} {532} (\bibinfo
  {year} {1981})},\ \bibinfo {note} {[Pisma Zh. Eksp. Teor.
  Fiz.33,549(1981)]}\BibitemShut {NoStop}%
\bibitem [{\citenamefont {Hawking}(1982)}]{Hawking:1982cz}%
  \BibitemOpen
  \bibfield  {author} {\bibinfo {author} {\bibfnamefont {S.~W.}\ \bibnamefont
  {Hawking}},\ }\href {\doibase 10.1016/0370-2693(82)90373-2} {\bibfield
  {journal} {\bibinfo  {journal} {Phys. Lett.}\ }\textbf {\bibinfo {volume}
  {B115}},\ \bibinfo {pages} {295} (\bibinfo {year} {1982})}\BibitemShut
  {NoStop}%
\bibitem [{\citenamefont {Starobinsky}(1982)}]{Starobinsky:1982ee}%
  \BibitemOpen
  \bibfield  {author} {\bibinfo {author} {\bibfnamefont {A.~A.}\ \bibnamefont
  {Starobinsky}},\ }\href {\doibase 10.1016/0370-2693(82)90541-X} {\bibfield
  {journal} {\bibinfo  {journal} {Phys. Lett.}\ }\textbf {\bibinfo {volume}
  {B117}},\ \bibinfo {pages} {175} (\bibinfo {year} {1982})}\BibitemShut
  {NoStop}%
\bibitem [{\citenamefont {Guth}\ and\ \citenamefont {Pi}(1982)}]{Guth:1982ec}%
  \BibitemOpen
  \bibfield  {author} {\bibinfo {author} {\bibfnamefont {A.~H.}\ \bibnamefont
  {Guth}}\ and\ \bibinfo {author} {\bibfnamefont {S.~Y.}\ \bibnamefont {Pi}},\
  }\href {\doibase 10.1103/PhysRevLett.49.1110} {\bibfield  {journal} {\bibinfo
   {journal} {Phys. Rev. Lett.}\ }\textbf {\bibinfo {volume} {49}},\ \bibinfo
  {pages} {1110} (\bibinfo {year} {1982})}\BibitemShut {NoStop}%
\bibitem [{\citenamefont {Bardeen}\ \emph {et~al.}(1983)\citenamefont
  {Bardeen}, \citenamefont {Steinhardt},\ and\ \citenamefont
  {Turner}}]{Bardeen:1983qw}%
  \BibitemOpen
  \bibfield  {author} {\bibinfo {author} {\bibfnamefont {J.~M.}\ \bibnamefont
  {Bardeen}}, \bibinfo {author} {\bibfnamefont {P.~J.}\ \bibnamefont
  {Steinhardt}}, \ and\ \bibinfo {author} {\bibfnamefont {M.~S.}\ \bibnamefont
  {Turner}},\ }\href {\doibase 10.1103/PhysRevD.28.679} {\bibfield  {journal}
  {\bibinfo  {journal} {Phys. Rev.}\ }\textbf {\bibinfo {volume} {D28}},\
  \bibinfo {pages} {679} (\bibinfo {year} {1983})}\BibitemShut {NoStop}%
\bibitem [{\citenamefont {Mollerach}(1990)}]{Mollerach:1989hu}%
  \BibitemOpen
  \bibfield  {author} {\bibinfo {author} {\bibfnamefont {S.}~\bibnamefont
  {Mollerach}},\ }\href {\doibase 10.1103/PhysRevD.42.313} {\bibfield
  {journal} {\bibinfo  {journal} {Phys. Rev.}\ }\textbf {\bibinfo {volume}
  {D42}},\ \bibinfo {pages} {313} (\bibinfo {year} {1990})}\BibitemShut
  {NoStop}%
\bibitem [{\citenamefont {Linde}\ and\ \citenamefont
  {Mukhanov}(1997)}]{Linde:1996gt}%
  \BibitemOpen
  \bibfield  {author} {\bibinfo {author} {\bibfnamefont {A.~D.}\ \bibnamefont
  {Linde}}\ and\ \bibinfo {author} {\bibfnamefont {V.~F.}\ \bibnamefont
  {Mukhanov}},\ }\href {\doibase 10.1103/PhysRevD.56.R535} {\bibfield
  {journal} {\bibinfo  {journal} {Phys. Rev.}\ }\textbf {\bibinfo {volume}
  {D56}},\ \bibinfo {pages} {535} (\bibinfo {year} {1997})},\ \Eprint
  {http://arxiv.org/abs/astro-ph/9610219} {arXiv:astro-ph/9610219 [astro-ph]}
  \BibitemShut {NoStop}%
\bibitem [{\citenamefont {Enqvist}\ and\ \citenamefont
  {Sloth}(2002)}]{Enqvist:2001zp}%
  \BibitemOpen
  \bibfield  {author} {\bibinfo {author} {\bibfnamefont {K.}~\bibnamefont
  {Enqvist}}\ and\ \bibinfo {author} {\bibfnamefont {M.~S.}\ \bibnamefont
  {Sloth}},\ }\href {\doibase 10.1016/S0550-3213(02)00043-3} {\bibfield
  {journal} {\bibinfo  {journal} {Nucl. Phys.}\ }\textbf {\bibinfo {volume}
  {B626}},\ \bibinfo {pages} {395} (\bibinfo {year} {2002})},\ \Eprint
  {http://arxiv.org/abs/hep-ph/0109214} {arXiv:hep-ph/0109214 [hep-ph]}
  \BibitemShut {NoStop}%
\bibitem [{\citenamefont {Lyth}\ and\ \citenamefont
  {Wands}(2002)}]{Lyth:2001nq}%
  \BibitemOpen
  \bibfield  {author} {\bibinfo {author} {\bibfnamefont {D.~H.}\ \bibnamefont
  {Lyth}}\ and\ \bibinfo {author} {\bibfnamefont {D.}~\bibnamefont {Wands}},\
  }\href {\doibase 10.1016/S0370-2693(01)01366-1} {\bibfield  {journal}
  {\bibinfo  {journal} {Phys. Lett.}\ }\textbf {\bibinfo {volume} {B524}},\
  \bibinfo {pages} {5} (\bibinfo {year} {2002})},\ \Eprint
  {http://arxiv.org/abs/hep-ph/0110002} {arXiv:hep-ph/0110002 [hep-ph]}
  \BibitemShut {NoStop}%
\bibitem [{\citenamefont {Moroi}\ and\ \citenamefont
  {Takahashi}(2001)}]{Moroi:2001ct}%
  \BibitemOpen
  \bibfield  {author} {\bibinfo {author} {\bibfnamefont {T.}~\bibnamefont
  {Moroi}}\ and\ \bibinfo {author} {\bibfnamefont {T.}~\bibnamefont
  {Takahashi}},\ }\href {\doibase 10.1016/S0370-2693(01)01295-3} {\bibfield
  {journal} {\bibinfo  {journal} {Phys. Lett.}\ }\textbf {\bibinfo {volume}
  {B522}},\ \bibinfo {pages} {215} (\bibinfo {year} {2001})},\ \bibinfo {note}
  {[Erratum: Phys. Lett.B539,303(2002)]},\ \Eprint
  {http://arxiv.org/abs/hep-ph/0110096} {arXiv:hep-ph/0110096 [hep-ph]}
  \BibitemShut {NoStop}%
\bibitem [{\citenamefont {Dvali}\ \emph {et~al.}(2004)\citenamefont {Dvali},
  \citenamefont {Gruzinov},\ and\ \citenamefont {Zaldarriaga}}]{Dvali:2003em}%
  \BibitemOpen
  \bibfield  {author} {\bibinfo {author} {\bibfnamefont {G.}~\bibnamefont
  {Dvali}}, \bibinfo {author} {\bibfnamefont {A.}~\bibnamefont {Gruzinov}}, \
  and\ \bibinfo {author} {\bibfnamefont {M.}~\bibnamefont {Zaldarriaga}},\
  }\href {\doibase 10.1103/PhysRevD.69.023505} {\bibfield  {journal} {\bibinfo
  {journal} {Phys. Rev.}\ }\textbf {\bibinfo {volume} {D69}},\ \bibinfo {pages}
  {023505} (\bibinfo {year} {2004})},\ \Eprint
  {http://arxiv.org/abs/astro-ph/0303591} {arXiv:astro-ph/0303591 [astro-ph]}
  \BibitemShut {NoStop}%
\bibitem [{\citenamefont {Kofman}(2003)}]{Kofman:2003nx}%
  \BibitemOpen
  \bibfield  {author} {\bibinfo {author} {\bibfnamefont {L.}~\bibnamefont
  {Kofman}},\ }\href@noop {} {\  (\bibinfo {year} {2003})},\ \Eprint
  {http://arxiv.org/abs/astro-ph/0303614} {arXiv:astro-ph/0303614 [astro-ph]}
  \BibitemShut {NoStop}%
\bibitem [{\citenamefont {Linde}(1983)}]{Linde:1983gd}%
  \BibitemOpen
  \bibfield  {author} {\bibinfo {author} {\bibfnamefont {A.~D.}\ \bibnamefont
  {Linde}},\ }\href {\doibase 10.1016/0370-2693(83)90837-7} {\bibfield
  {journal} {\bibinfo  {journal} {Phys. Lett.}\ }\textbf {\bibinfo {volume}
  {B129}},\ \bibinfo {pages} {177} (\bibinfo {year} {1983})}\BibitemShut
  {NoStop}%
\bibitem [{\citenamefont {Linde}(1990)}]{Linde:2005ht}%
  \BibitemOpen
  \bibfield  {author} {\bibinfo {author} {\bibfnamefont {A.~D.}\ \bibnamefont
  {Linde}},\ }\href@noop {} {\bibfield  {journal} {\bibinfo  {journal}
  {Contemp. Concepts Phys.}\ }\textbf {\bibinfo {volume} {5}},\ \bibinfo
  {pages} {1} (\bibinfo {year} {1990})},\ \Eprint
  {http://arxiv.org/abs/hep-th/0503203} {arXiv:hep-th/0503203 [hep-th]}
  \BibitemShut {NoStop}%
\bibitem [{\citenamefont {Ade}\ \emph {et~al.}(2015{\natexlab{a}})\citenamefont
  {Ade} \emph {et~al.}}]{Ade:2015lrj}%
  \BibitemOpen
  \bibfield  {author} {\bibinfo {author} {\bibfnamefont {P.~A.~R.}\
  \bibnamefont {Ade}} \emph {et~al.} (\bibinfo {collaboration} {Planck}),\
  }\href@noop {} {\  (\bibinfo {year} {2015}{\natexlab{a}})},\ \Eprint
  {http://arxiv.org/abs/1502.02114} {arXiv:1502.02114 [astro-ph.CO]}
  \BibitemShut {NoStop}%
\bibitem [{\citenamefont {Ovrut}\ and\ \citenamefont
  {Steinhardt}(1983)}]{Ovrut:1983my}%
  \BibitemOpen
  \bibfield  {author} {\bibinfo {author} {\bibfnamefont {B.~A.}\ \bibnamefont
  {Ovrut}}\ and\ \bibinfo {author} {\bibfnamefont {P.~J.}\ \bibnamefont
  {Steinhardt}},\ }\href {\doibase 10.1016/0370-2693(83)90551-8} {\bibfield
  {journal} {\bibinfo  {journal} {Phys. Lett.}\ }\textbf {\bibinfo {volume}
  {B133}},\ \bibinfo {pages} {161} (\bibinfo {year} {1983})}\BibitemShut
  {NoStop}%
\bibitem [{\citenamefont {Holman}\ \emph {et~al.}(1984)\citenamefont {Holman},
  \citenamefont {Ramond},\ and\ \citenamefont {Ross}}]{Holman:1984yj}%
  \BibitemOpen
  \bibfield  {author} {\bibinfo {author} {\bibfnamefont {R.}~\bibnamefont
  {Holman}}, \bibinfo {author} {\bibfnamefont {P.}~\bibnamefont {Ramond}}, \
  and\ \bibinfo {author} {\bibfnamefont {G.~G.}\ \bibnamefont {Ross}},\ }\href
  {\doibase 10.1016/0370-2693(84)91729-5} {\bibfield  {journal} {\bibinfo
  {journal} {Phys. Lett.}\ }\textbf {\bibinfo {volume} {B137}},\ \bibinfo
  {pages} {343} (\bibinfo {year} {1984})}\BibitemShut {NoStop}%
\bibitem [{\citenamefont {Goncharov}\ and\ \citenamefont
  {Linde}(1984)}]{Goncharov:1983mw}%
  \BibitemOpen
  \bibfield  {author} {\bibinfo {author} {\bibfnamefont {A.~B.}\ \bibnamefont
  {Goncharov}}\ and\ \bibinfo {author} {\bibfnamefont {A.~D.}\ \bibnamefont
  {Linde}},\ }\href {\doibase 10.1016/0370-2693(84)90027-3} {\bibfield
  {journal} {\bibinfo  {journal} {Phys. Lett.}\ }\textbf {\bibinfo {volume}
  {B139}},\ \bibinfo {pages} {27} (\bibinfo {year} {1984})}\BibitemShut
  {NoStop}%
\bibitem [{\citenamefont {Coughlan}\ \emph {et~al.}(1984)\citenamefont
  {Coughlan}, \citenamefont {Holman}, \citenamefont {Ramond},\ and\
  \citenamefont {Ross}}]{Coughlan:1984yk}%
  \BibitemOpen
  \bibfield  {author} {\bibinfo {author} {\bibfnamefont {G.~D.}\ \bibnamefont
  {Coughlan}}, \bibinfo {author} {\bibfnamefont {R.}~\bibnamefont {Holman}},
  \bibinfo {author} {\bibfnamefont {P.}~\bibnamefont {Ramond}}, \ and\ \bibinfo
  {author} {\bibfnamefont {G.~G.}\ \bibnamefont {Ross}},\ }\href {\doibase
  10.1016/0370-2693(84)91043-8} {\bibfield  {journal} {\bibinfo  {journal}
  {Phys. Lett.}\ }\textbf {\bibinfo {volume} {B140}},\ \bibinfo {pages} {44}
  (\bibinfo {year} {1984})}\BibitemShut {NoStop}%
\bibitem [{\citenamefont {McDonald}(2003)}]{McDonald:2003xq}%
  \BibitemOpen
  \bibfield  {author} {\bibinfo {author} {\bibfnamefont {J.}~\bibnamefont
  {McDonald}},\ }\href {\doibase 10.1103/PhysRevD.68.043505} {\bibfield
  {journal} {\bibinfo  {journal} {Phys. Rev.}\ }\textbf {\bibinfo {volume}
  {D68}},\ \bibinfo {pages} {043505} (\bibinfo {year} {2003})},\ \Eprint
  {http://arxiv.org/abs/hep-ph/0302222} {arXiv:hep-ph/0302222 [hep-ph]}
  \BibitemShut {NoStop}%
\bibitem [{\citenamefont {McDonald}(2004)}]{McDonald:2004by}%
  \BibitemOpen
  \bibfield  {author} {\bibinfo {author} {\bibfnamefont {J.}~\bibnamefont
  {McDonald}},\ }\href {\doibase 10.1103/PhysRevD.70.063520} {\bibfield
  {journal} {\bibinfo  {journal} {Phys. Rev.}\ }\textbf {\bibinfo {volume}
  {D70}},\ \bibinfo {pages} {063520} (\bibinfo {year} {2004})},\ \Eprint
  {http://arxiv.org/abs/hep-ph/0404154} {arXiv:hep-ph/0404154 [hep-ph]}
  \BibitemShut {NoStop}%
\bibitem [{\citenamefont {Kawasaki}\ \emph {et~al.}(2011)\citenamefont
  {Kawasaki}, \citenamefont {Kobayashi},\ and\ \citenamefont
  {Takahashi}}]{Kawasaki:2011pd}%
  \BibitemOpen
  \bibfield  {author} {\bibinfo {author} {\bibfnamefont {M.}~\bibnamefont
  {Kawasaki}}, \bibinfo {author} {\bibfnamefont {T.}~\bibnamefont {Kobayashi}},
  \ and\ \bibinfo {author} {\bibfnamefont {F.}~\bibnamefont {Takahashi}},\
  }\href {\doibase 10.1103/PhysRevD.84.123506, 10.1103/PhysRevD.85.029905}
  {\bibfield  {journal} {\bibinfo  {journal} {Phys. Rev.}\ }\textbf {\bibinfo
  {volume} {D84}},\ \bibinfo {pages} {123506} (\bibinfo {year} {2011})},\
  \Eprint {http://arxiv.org/abs/1107.6011} {arXiv:1107.6011 [astro-ph.CO]}
  \BibitemShut {NoStop}%
\bibitem [{\citenamefont {Lyth}\ \emph {et~al.}(2003)\citenamefont {Lyth},
  \citenamefont {Ungarelli},\ and\ \citenamefont {Wands}}]{Lyth:2002my}%
  \BibitemOpen
  \bibfield  {author} {\bibinfo {author} {\bibfnamefont {D.~H.}\ \bibnamefont
  {Lyth}}, \bibinfo {author} {\bibfnamefont {C.}~\bibnamefont {Ungarelli}}, \
  and\ \bibinfo {author} {\bibfnamefont {D.}~\bibnamefont {Wands}},\ }\href
  {\doibase 10.1103/PhysRevD.67.023503} {\bibfield  {journal} {\bibinfo
  {journal} {Phys. Rev.}\ }\textbf {\bibinfo {volume} {D67}},\ \bibinfo {pages}
  {023503} (\bibinfo {year} {2003})},\ \Eprint
  {http://arxiv.org/abs/astro-ph/0208055} {arXiv:astro-ph/0208055 [astro-ph]}
  \BibitemShut {NoStop}%
\bibitem [{\citenamefont {Lyth}\ and\ \citenamefont
  {Wands}(2003)}]{Lyth:2003ip}%
  \BibitemOpen
  \bibfield  {author} {\bibinfo {author} {\bibfnamefont {D.~H.}\ \bibnamefont
  {Lyth}}\ and\ \bibinfo {author} {\bibfnamefont {D.}~\bibnamefont {Wands}},\
  }\href {\doibase 10.1103/PhysRevD.68.103516} {\bibfield  {journal} {\bibinfo
  {journal} {Phys. Rev.}\ }\textbf {\bibinfo {volume} {D68}},\ \bibinfo {pages}
  {103516} (\bibinfo {year} {2003})},\ \Eprint
  {http://arxiv.org/abs/astro-ph/0306500} {arXiv:astro-ph/0306500 [astro-ph]}
  \BibitemShut {NoStop}%
\bibitem [{\citenamefont {Enqvist}\ and\ \citenamefont
  {Takahashi}(2013)}]{Enqvist:2013paa}%
  \BibitemOpen
  \bibfield  {author} {\bibinfo {author} {\bibfnamefont {K.}~\bibnamefont
  {Enqvist}}\ and\ \bibinfo {author} {\bibfnamefont {T.}~\bibnamefont
  {Takahashi}},\ }\href {\doibase 10.1088/1475-7516/2013/10/034} {\bibfield
  {journal} {\bibinfo  {journal} {JCAP}\ }\textbf {\bibinfo {volume} {1310}},\
  \bibinfo {pages} {034} (\bibinfo {year} {2013})},\ \Eprint
  {http://arxiv.org/abs/1306.5958} {arXiv:1306.5958 [astro-ph.CO]} \BibitemShut
  {NoStop}%
\bibitem [{\citenamefont {Fujita}\ \emph
  {et~al.}(2014{\natexlab{a}})\citenamefont {Fujita}, \citenamefont
  {Kawasaki},\ and\ \citenamefont {Yokoyama}}]{Fujita:2014iaa}%
  \BibitemOpen
  \bibfield  {author} {\bibinfo {author} {\bibfnamefont {T.}~\bibnamefont
  {Fujita}}, \bibinfo {author} {\bibfnamefont {M.}~\bibnamefont {Kawasaki}}, \
  and\ \bibinfo {author} {\bibfnamefont {S.}~\bibnamefont {Yokoyama}},\ }\href
  {\doibase 10.1088/1475-7516/2014/09/015} {\bibfield  {journal} {\bibinfo
  {journal} {JCAP}\ }\textbf {\bibinfo {volume} {1409}},\ \bibinfo {pages}
  {015} (\bibinfo {year} {2014}{\natexlab{a}})},\ \Eprint
  {http://arxiv.org/abs/1404.0951} {arXiv:1404.0951 [astro-ph.CO]} \BibitemShut
  {NoStop}%
\bibitem [{\citenamefont {Ikegami}\ and\ \citenamefont
  {Moroi}(2004)}]{Ikegami:2004ve}%
  \BibitemOpen
  \bibfield  {author} {\bibinfo {author} {\bibfnamefont {M.}~\bibnamefont
  {Ikegami}}\ and\ \bibinfo {author} {\bibfnamefont {T.}~\bibnamefont
  {Moroi}},\ }\href {\doibase 10.1103/PhysRevD.70.083515} {\bibfield  {journal}
  {\bibinfo  {journal} {Phys. Rev.}\ }\textbf {\bibinfo {volume} {D70}},\
  \bibinfo {pages} {083515} (\bibinfo {year} {2004})},\ \Eprint
  {http://arxiv.org/abs/hep-ph/0404253} {arXiv:hep-ph/0404253 [hep-ph]}
  \BibitemShut {NoStop}%
\bibitem [{\citenamefont {Gordon}\ and\ \citenamefont
  {Lewis}(2003)}]{Gordon:2002gv}%
  \BibitemOpen
  \bibfield  {author} {\bibinfo {author} {\bibfnamefont {C.}~\bibnamefont
  {Gordon}}\ and\ \bibinfo {author} {\bibfnamefont {A.}~\bibnamefont {Lewis}},\
  }\href {\doibase 10.1103/PhysRevD.67.123513} {\bibfield  {journal} {\bibinfo
  {journal} {Phys. Rev.}\ }\textbf {\bibinfo {volume} {D67}},\ \bibinfo {pages}
  {123513} (\bibinfo {year} {2003})},\ \Eprint
  {http://arxiv.org/abs/astro-ph/0212248} {arXiv:astro-ph/0212248 [astro-ph]}
  \BibitemShut {NoStop}%
\bibitem [{\citenamefont {Harigaya}\ \emph {et~al.}(2013)\citenamefont
  {Harigaya}, \citenamefont {Ibe}, \citenamefont {Kawasaki},\ and\
  \citenamefont {Yanagida}}]{Harigaya:2012up}%
  \BibitemOpen
  \bibfield  {author} {\bibinfo {author} {\bibfnamefont {K.}~\bibnamefont
  {Harigaya}}, \bibinfo {author} {\bibfnamefont {M.}~\bibnamefont {Ibe}},
  \bibinfo {author} {\bibfnamefont {M.}~\bibnamefont {Kawasaki}}, \ and\
  \bibinfo {author} {\bibfnamefont {T.~T.}\ \bibnamefont {Yanagida}},\ }\href
  {\doibase 10.1103/PhysRevD.87.063514} {\bibfield  {journal} {\bibinfo
  {journal} {Phys. Rev.}\ }\textbf {\bibinfo {volume} {D87}},\ \bibinfo {pages}
  {063514} (\bibinfo {year} {2013})},\ \Eprint {http://arxiv.org/abs/1211.3535}
  {arXiv:1211.3535 [hep-ph]} \BibitemShut {NoStop}%
\bibitem [{\citenamefont {Harigaya}\ \emph {et~al.}(2014)\citenamefont
  {Harigaya}, \citenamefont {Hayakawa}, \citenamefont {Kawasaki},\ and\
  \citenamefont {Yokoyama}}]{Harigaya:2014bsa}%
  \BibitemOpen
  \bibfield  {author} {\bibinfo {author} {\bibfnamefont {K.}~\bibnamefont
  {Harigaya}}, \bibinfo {author} {\bibfnamefont {T.}~\bibnamefont {Hayakawa}},
  \bibinfo {author} {\bibfnamefont {M.}~\bibnamefont {Kawasaki}}, \ and\
  \bibinfo {author} {\bibfnamefont {S.}~\bibnamefont {Yokoyama}},\ }\href
  {\doibase 10.1088/1475-7516/2014/10/068} {\bibfield  {journal} {\bibinfo
  {journal} {JCAP}\ }\textbf {\bibinfo {volume} {1410}},\ \bibinfo {pages}
  {068} (\bibinfo {year} {2014})},\ \Eprint {http://arxiv.org/abs/1409.1669}
  {arXiv:1409.1669 [hep-ph]} \BibitemShut {NoStop}%
\bibitem [{\citenamefont {Fukugita}\ and\ \citenamefont
  {Yanagida}(1986)}]{Fukugita:1986hr}%
  \BibitemOpen
  \bibfield  {author} {\bibinfo {author} {\bibfnamefont {M.}~\bibnamefont
  {Fukugita}}\ and\ \bibinfo {author} {\bibfnamefont {T.}~\bibnamefont
  {Yanagida}},\ }\href {\doibase 10.1016/0370-2693(86)91126-3} {\bibfield
  {journal} {\bibinfo  {journal} {Phys. Lett.}\ }\textbf {\bibinfo {volume}
  {B174}},\ \bibinfo {pages} {45} (\bibinfo {year} {1986})}\BibitemShut
  {NoStop}%
\bibitem [{\citenamefont {Murayama}\ and\ \citenamefont
  {Yanagida}(1994)}]{Murayama:1993em}%
  \BibitemOpen
  \bibfield  {author} {\bibinfo {author} {\bibfnamefont {H.}~\bibnamefont
  {Murayama}}\ and\ \bibinfo {author} {\bibfnamefont {T.}~\bibnamefont
  {Yanagida}},\ }\href {\doibase 10.1016/0370-2693(94)91164-9} {\bibfield
  {journal} {\bibinfo  {journal} {Phys. Lett.}\ }\textbf {\bibinfo {volume}
  {B322}},\ \bibinfo {pages} {349} (\bibinfo {year} {1994})},\ \Eprint
  {http://arxiv.org/abs/hep-ph/9310297} {arXiv:hep-ph/9310297 [hep-ph]}
  \BibitemShut {NoStop}%
\bibitem [{\citenamefont {Moroi}\ and\ \citenamefont
  {Murayama}(2003)}]{Moroi:2002vx}%
  \BibitemOpen
  \bibfield  {author} {\bibinfo {author} {\bibfnamefont {T.}~\bibnamefont
  {Moroi}}\ and\ \bibinfo {author} {\bibfnamefont {H.}~\bibnamefont
  {Murayama}},\ }\href {\doibase 10.1016/S0370-2693(02)03227-6} {\bibfield
  {journal} {\bibinfo  {journal} {Phys. Lett.}\ }\textbf {\bibinfo {volume}
  {B553}},\ \bibinfo {pages} {126} (\bibinfo {year} {2003})},\ \Eprint
  {http://arxiv.org/abs/hep-ph/0211019} {arXiv:hep-ph/0211019 [hep-ph]}
  \BibitemShut {NoStop}%
\bibitem [{\citenamefont {Postma}(2003)}]{Postma:2002et}%
  \BibitemOpen
  \bibfield  {author} {\bibinfo {author} {\bibfnamefont {M.}~\bibnamefont
  {Postma}},\ }\href {\doibase 10.1103/PhysRevD.67.063518} {\bibfield
  {journal} {\bibinfo  {journal} {Phys. Rev.}\ }\textbf {\bibinfo {volume}
  {D67}},\ \bibinfo {pages} {063518} (\bibinfo {year} {2003})},\ \Eprint
  {http://arxiv.org/abs/hep-ph/0212005} {arXiv:hep-ph/0212005 [hep-ph]}
  \BibitemShut {NoStop}%
\bibitem [{\citenamefont {Affleck}\ and\ \citenamefont
  {Dine}(1985)}]{Affleck:1984fy}%
  \BibitemOpen
  \bibfield  {author} {\bibinfo {author} {\bibfnamefont {I.}~\bibnamefont
  {Affleck}}\ and\ \bibinfo {author} {\bibfnamefont {M.}~\bibnamefont {Dine}},\
  }\href {\doibase 10.1016/0550-3213(85)90021-5} {\bibfield  {journal}
  {\bibinfo  {journal} {Nucl. Phys.}\ }\textbf {\bibinfo {volume} {B249}},\
  \bibinfo {pages} {361} (\bibinfo {year} {1985})}\BibitemShut {NoStop}%
\bibitem [{\citenamefont {Dine}\ \emph {et~al.}(1996)\citenamefont {Dine},
  \citenamefont {Randall},\ and\ \citenamefont {Thomas}}]{Dine:1995kz}%
  \BibitemOpen
  \bibfield  {author} {\bibinfo {author} {\bibfnamefont {M.}~\bibnamefont
  {Dine}}, \bibinfo {author} {\bibfnamefont {L.}~\bibnamefont {Randall}}, \
  and\ \bibinfo {author} {\bibfnamefont {S.~D.}\ \bibnamefont {Thomas}},\
  }\href {\doibase 10.1016/0550-3213(95)00538-2} {\bibfield  {journal}
  {\bibinfo  {journal} {Nucl. Phys.}\ }\textbf {\bibinfo {volume} {B458}},\
  \bibinfo {pages} {291} (\bibinfo {year} {1996})},\ \Eprint
  {http://arxiv.org/abs/hep-ph/9507453} {arXiv:hep-ph/9507453 [hep-ph]}
  \BibitemShut {NoStop}%
\bibitem [{\citenamefont {Buchmuller}\ \emph {et~al.}(2005)\citenamefont
  {Buchmuller}, \citenamefont {Di~Bari},\ and\ \citenamefont
  {Plumacher}}]{Buchmuller:2004nz}%
  \BibitemOpen
  \bibfield  {author} {\bibinfo {author} {\bibfnamefont {W.}~\bibnamefont
  {Buchmuller}}, \bibinfo {author} {\bibfnamefont {P.}~\bibnamefont {Di~Bari}},
  \ and\ \bibinfo {author} {\bibfnamefont {M.}~\bibnamefont {Plumacher}},\
  }\href {\doibase 10.1016/j.aop.2004.02.003} {\bibfield  {journal} {\bibinfo
  {journal} {Annals Phys.}\ }\textbf {\bibinfo {volume} {315}},\ \bibinfo
  {pages} {305} (\bibinfo {year} {2005})},\ \Eprint
  {http://arxiv.org/abs/hep-ph/0401240} {arXiv:hep-ph/0401240 [hep-ph]}
  \BibitemShut {NoStop}%
\bibitem [{\citenamefont {Hamaguchi}\ \emph {et~al.}(2002)\citenamefont
  {Hamaguchi}, \citenamefont {Murayama},\ and\ \citenamefont
  {Yanagida}}]{Hamaguchi:2001gw}%
  \BibitemOpen
  \bibfield  {author} {\bibinfo {author} {\bibfnamefont {K.}~\bibnamefont
  {Hamaguchi}}, \bibinfo {author} {\bibfnamefont {H.}~\bibnamefont {Murayama}},
  \ and\ \bibinfo {author} {\bibfnamefont {T.}~\bibnamefont {Yanagida}},\
  }\href {\doibase 10.1103/PhysRevD.65.043512} {\bibfield  {journal} {\bibinfo
  {journal} {Phys. Rev.}\ }\textbf {\bibinfo {volume} {D65}},\ \bibinfo {pages}
  {043512} (\bibinfo {year} {2002})},\ \Eprint
  {http://arxiv.org/abs/hep-ph/0109030} {arXiv:hep-ph/0109030 [hep-ph]}
  \BibitemShut {NoStop}%
\bibitem [{\citenamefont {Sasaki}\ and\ \citenamefont
  {Stewart}(1996)}]{Sasaki:1995aw}%
  \BibitemOpen
  \bibfield  {author} {\bibinfo {author} {\bibfnamefont {M.}~\bibnamefont
  {Sasaki}}\ and\ \bibinfo {author} {\bibfnamefont {E.~D.}\ \bibnamefont
  {Stewart}},\ }\href {\doibase 10.1143/PTP.95.71} {\bibfield  {journal}
  {\bibinfo  {journal} {Prog. Theor. Phys.}\ }\textbf {\bibinfo {volume}
  {95}},\ \bibinfo {pages} {71} (\bibinfo {year} {1996})},\ \Eprint
  {http://arxiv.org/abs/astro-ph/9507001} {arXiv:astro-ph/9507001 [astro-ph]}
  \BibitemShut {NoStop}%
\bibitem [{\citenamefont {Wands}\ \emph {et~al.}(2000)\citenamefont {Wands},
  \citenamefont {Malik}, \citenamefont {Lyth},\ and\ \citenamefont
  {Liddle}}]{Wands:2000dp}%
  \BibitemOpen
  \bibfield  {author} {\bibinfo {author} {\bibfnamefont {D.}~\bibnamefont
  {Wands}}, \bibinfo {author} {\bibfnamefont {K.~A.}\ \bibnamefont {Malik}},
  \bibinfo {author} {\bibfnamefont {D.~H.}\ \bibnamefont {Lyth}}, \ and\
  \bibinfo {author} {\bibfnamefont {A.~R.}\ \bibnamefont {Liddle}},\ }\href
  {\doibase 10.1103/PhysRevD.62.043527} {\bibfield  {journal} {\bibinfo
  {journal} {Phys. Rev.}\ }\textbf {\bibinfo {volume} {D62}},\ \bibinfo {pages}
  {043527} (\bibinfo {year} {2000})},\ \Eprint
  {http://arxiv.org/abs/astro-ph/0003278} {arXiv:astro-ph/0003278 [astro-ph]}
  \BibitemShut {NoStop}%
\bibitem [{\citenamefont {Lyth}\ \emph {et~al.}(2005)\citenamefont {Lyth},
  \citenamefont {Malik},\ and\ \citenamefont {Sasaki}}]{Lyth:2004gb}%
  \BibitemOpen
  \bibfield  {author} {\bibinfo {author} {\bibfnamefont {D.~H.}\ \bibnamefont
  {Lyth}}, \bibinfo {author} {\bibfnamefont {K.~A.}\ \bibnamefont {Malik}}, \
  and\ \bibinfo {author} {\bibfnamefont {M.}~\bibnamefont {Sasaki}},\ }\href
  {\doibase 10.1088/1475-7516/2005/05/004} {\bibfield  {journal} {\bibinfo
  {journal} {JCAP}\ }\textbf {\bibinfo {volume} {0505}},\ \bibinfo {pages}
  {004} (\bibinfo {year} {2005})},\ \Eprint
  {http://arxiv.org/abs/astro-ph/0411220} {arXiv:astro-ph/0411220 [astro-ph]}
  \BibitemShut {NoStop}%
\bibitem [{\citenamefont {Kawasaki}\ and\ \citenamefont
  {Takesako}(2012{\natexlab{a}})}]{Kawasaki:2011zi}%
  \BibitemOpen
  \bibfield  {author} {\bibinfo {author} {\bibfnamefont {M.}~\bibnamefont
  {Kawasaki}}\ and\ \bibinfo {author} {\bibfnamefont {T.}~\bibnamefont
  {Takesako}},\ }\href {\doibase 10.1016/j.physletb.2012.03.069} {\bibfield
  {journal} {\bibinfo  {journal} {Phys. Lett.}\ }\textbf {\bibinfo {volume}
  {B711}},\ \bibinfo {pages} {173} (\bibinfo {year} {2012}{\natexlab{a}})},\
  \Eprint {http://arxiv.org/abs/1112.5823} {arXiv:1112.5823 [hep-ph]}
  \BibitemShut {NoStop}%
\bibitem [{\citenamefont {Kawasaki}\ and\ \citenamefont
  {Takesako}(2012{\natexlab{b}})}]{Kawasaki:2012qm}%
  \BibitemOpen
  \bibfield  {author} {\bibinfo {author} {\bibfnamefont {M.}~\bibnamefont
  {Kawasaki}}\ and\ \bibinfo {author} {\bibfnamefont {T.}~\bibnamefont
  {Takesako}},\ }\href {\doibase 10.1016/j.physletb.2012.10.080} {\bibfield
  {journal} {\bibinfo  {journal} {Phys. Lett.}\ }\textbf {\bibinfo {volume}
  {B718}},\ \bibinfo {pages} {522} (\bibinfo {year} {2012}{\natexlab{b}})},\
  \Eprint {http://arxiv.org/abs/1208.1323} {arXiv:1208.1323 [hep-ph]}
  \BibitemShut {NoStop}%
\bibitem [{\citenamefont {Kawasaki}\ \emph {et~al.}(2013)\citenamefont
  {Kawasaki}, \citenamefont {Takahashi},\ and\ \citenamefont
  {Takesako}}]{Kawasaki:2012rs}%
  \BibitemOpen
  \bibfield  {author} {\bibinfo {author} {\bibfnamefont {M.}~\bibnamefont
  {Kawasaki}}, \bibinfo {author} {\bibfnamefont {F.}~\bibnamefont {Takahashi}},
  \ and\ \bibinfo {author} {\bibfnamefont {T.}~\bibnamefont {Takesako}},\
  }\href {\doibase 10.1088/1475-7516/2013/04/008} {\bibfield  {journal}
  {\bibinfo  {journal} {JCAP}\ }\textbf {\bibinfo {volume} {1304}},\ \bibinfo
  {pages} {008} (\bibinfo {year} {2013})},\ \Eprint
  {http://arxiv.org/abs/1211.4921} {arXiv:1211.4921 [hep-ph]} \BibitemShut
  {NoStop}%
\bibitem [{\citenamefont {Fujita}\ \emph {et~al.}(2013)\citenamefont {Fujita},
  \citenamefont {Harigaya},\ and\ \citenamefont {Kawasaki}}]{Fujita:2013bka}%
  \BibitemOpen
  \bibfield  {author} {\bibinfo {author} {\bibfnamefont {T.}~\bibnamefont
  {Fujita}}, \bibinfo {author} {\bibfnamefont {K.}~\bibnamefont {Harigaya}}, \
  and\ \bibinfo {author} {\bibfnamefont {M.}~\bibnamefont {Kawasaki}},\ }\href
  {\doibase 10.1103/PhysRevD.88.123519} {\bibfield  {journal} {\bibinfo
  {journal} {Phys. Rev.}\ }\textbf {\bibinfo {volume} {D88}},\ \bibinfo {pages}
  {123519} (\bibinfo {year} {2013})},\ \Eprint {http://arxiv.org/abs/1306.6437}
  {arXiv:1306.6437 [astro-ph.CO]} \BibitemShut {NoStop}%
\bibitem [{\citenamefont {Fujita}\ \emph
  {et~al.}(2014{\natexlab{b}})\citenamefont {Fujita}, \citenamefont {Kawasaki},
  \citenamefont {Harigaya},\ and\ \citenamefont {Matsuda}}]{Fujita:2014hha}%
  \BibitemOpen
  \bibfield  {author} {\bibinfo {author} {\bibfnamefont {T.}~\bibnamefont
  {Fujita}}, \bibinfo {author} {\bibfnamefont {M.}~\bibnamefont {Kawasaki}},
  \bibinfo {author} {\bibfnamefont {K.}~\bibnamefont {Harigaya}}, \ and\
  \bibinfo {author} {\bibfnamefont {R.}~\bibnamefont {Matsuda}},\ }\href
  {\doibase 10.1103/PhysRevD.89.103501} {\bibfield  {journal} {\bibinfo
  {journal} {Phys. Rev.}\ }\textbf {\bibinfo {volume} {D89}},\ \bibinfo {pages}
  {103501} (\bibinfo {year} {2014}{\natexlab{b}})},\ \Eprint
  {http://arxiv.org/abs/1401.1909} {arXiv:1401.1909 [astro-ph.CO]} \BibitemShut
  {NoStop}%
\bibitem [{\citenamefont {Ade}\ \emph {et~al.}(2015{\natexlab{b}})\citenamefont
  {Ade} \emph {et~al.}}]{Ade:2015ava}%
  \BibitemOpen
  \bibfield  {author} {\bibinfo {author} {\bibfnamefont {P.~A.~R.}\
  \bibnamefont {Ade}} \emph {et~al.} (\bibinfo {collaboration} {Planck}),\
  }\href@noop {} {\  (\bibinfo {year} {2015}{\natexlab{b}})},\ \Eprint
  {http://arxiv.org/abs/1502.01592} {arXiv:1502.01592 [astro-ph.CO]}
  \BibitemShut {NoStop}%
\end{thebibliography}%

\end{document}